\documentclass[%
 aip,
 amsmath,amssymb,
 reprint,%
 onecolumn
]{revtex4-1}

\usepackage{graphicx}
\usepackage{dcolumn}
\usepackage{bm}
\usepackage{subcaption}
\usepackage{changepage}
\usepackage[utf8]{inputenc}
\usepackage[T1]{fontenc}
\usepackage{mathptmx}
\usepackage{tabularx}

\begin{document}

\preprint{AIP/123-QED}

\title[]{Acoustic characteristics of supersonic planar impinging jets}

\author{Nitish Arya}
\author{Ashoke De}%
 \email{ashoke@iitk.ac.in}
\affiliation{Indian Institute of Technology, Kanpur
}%


\begin{abstract}
	
The present work aims to study the tonal and broadband noise associated with a supersonic planar jet impinging on a flat plate. Five different cases are considered corresponding to different plate distance and angle of impingement. The near-field noise is analyzed using Large Eddy Simulation (LES). For this, we employ a low dispersive and dissipative Finite Volume solver using AUSM for inviscid fluxes and third order Runge-Kutta method for temporal discretization. The far-field noise is calculated using a hybrid approach in which the acoustic sources and mean flow are provided by the fluid solver while the far-field acoustic pressure is calculated by an acoustic solver. It solves Acoustic Perturbation Equations using a spectral/hp element method with a Discontinuous Galerkin projection. The present work highlights many important aspects of impinging jets. Firstly, the far-field sound with the effect of non-uniform base-fields is obtained which is important as the far-field acoustic pressure in most of the literature is reported either without any base-fields or by the application of uniform base-fields. Secondly, a correlation between the different modes of oscillations of the impinging jet and the production of impinging tones is studied using Modal Decomposition techniques. This analysis also sheds some light on the number of concurrent cycles for different nozzle to plate distance. Lastly, the relation between the phase lag term and the distance between the plate and the nozzle exit is highlighted using a vortex tracking method. This observation can serve to provide a better understanding of the sound generation process.

\end{abstract}

\maketitle

\begin{quotation}
	
\end{quotation}

\section{\label{sec:level1}Introduction}

The number of rocket launches has been steadily increasing every year. With the advent of private players in the space sector, there has been tremendous investment in space applications like internet services, navigation and surveillance, space exploration, space tourism, weather, and climate monitoring. Consequently, in the next few years, the number of rocket launches is bound to increase further. The noise generated due to the rocket launches is a major cause of stress among humans in urban areas and also detrimental to other life forms. Therefore, noise mitigation is a critical consideration for the rocket industry. A representative case of the rocket during the launch is a supersonic jet impinging on a surface. The present work aims to study the flow and acoustic properties of a supersonic planar jet impinging on a flat plate.

\begin{figure}
	\includegraphics[scale=0.5]{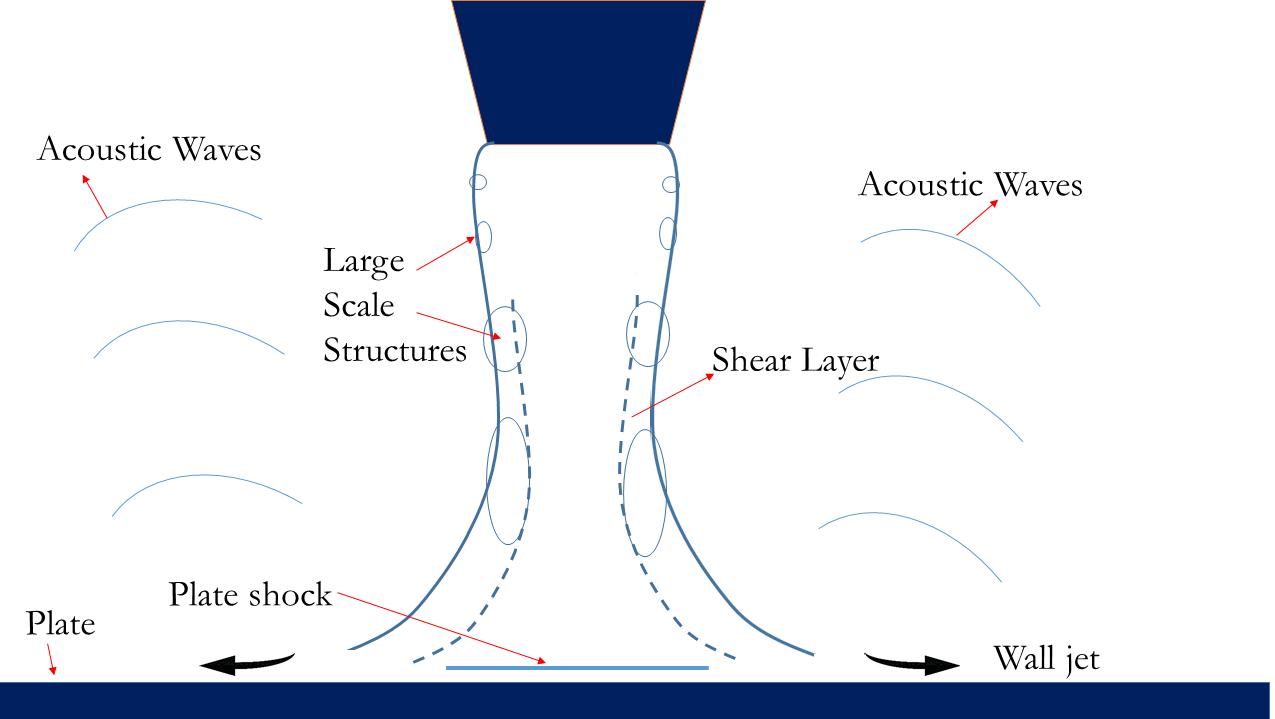}
	\centering
	\caption{\label{fig:1}Schematics of jet impingement process}	
\end{figure}

The schematics of an ideally expanded supersonic jet impingement process are presented in Fig. \ref{fig:1}. The large scale structures in the shear layer impinge on the plate giving rise to a wall jet that flows parallel to the wall. Since the flow before impinging is supersonic, a shock is formed near the plate which is referred to as plate shock, also called the stand-off shock. The impinging process is also characterized by the formation of acoustic waves that travel upstream. The acoustic waves excite the shear layer instabilities generating large scale structures that travel downstream and impinging on the plate, thus forming a feedback loop \cite{powell1988sound}. This feedback mechanism is similar to that found in edge tones \cite{powell1953edge,powell1961edgetone}and jet screech found in imperfectly expanded supersonic jets \cite{powell1953noise,powell1953mechanism}.
The analytical expression for the prediction of frequency, $f$, of the tones due to the jet impingement was provided by Powell \cite{powell1953noise,powell1953mechanism} which was similar to that for a supersonic jet undergoing screech.

\begin{equation}
	\frac{N+p}{f} = \int_H\frac{dh}{u_c} + \frac{h_A}{U_A}
\end{equation}   

Here, H is the distance of the plate from the nozzle, $u_c$ is the convective velocity of the large scale structures, $U_A$ is the speed with which the acoustic waves travel upstream covering a distance of $h_A$. $N$ is the number of concurrent cycles or the number of new disturbances formed in one cycle of feedback loop. The phase lag term $p$ i related to the delays associated with:

  1. The arrival of upstream wave at the lip and excitation of instabilities at the shear layer and-
  
  2. The impingement of the large-scale structures at the plate and the generation of upstream traveling acoustic waves
  
It has been found by Mitchell et al. \cite{mitchell2012visualization} that the time delay at the nozzle lip does not contribute much to the total lag.

The existence of impinging tones was first attributed to the feedback process between the nozzle and the plate by Ho and Nossier \cite{ho1981dynamics}. The authors made this observation for high subsonic jets. Powell \cite{powell1988sound} studied imperfectly expanded impinging jets for different plate sizes. He observed that when the plate size was comparable to the nozzle exit diameter, different tones were produced (or not produced) when the plate was kept at different places in the shock cell structure. He suggested that the tones were produced to the oscillation of the stand-off shock only when the plate was kept in the expansion region of the shock cell structure. For plate size much larger as compared to the exit diameter of the nozzle, he suggested that the source of the acoustic waves must lie on the plate itself. The existence of multiple tones for the impingement process had been confirmed by Krothapalli \cite{krothapalli1985discrete} for rectangular jets and Henderson et al. \cite{henderson1993experiments} for axisymmetric jets. Both the studies also reported a staging behavior dependent mainly on the plate distance, and to some extent, on the pressure ratio. Henderson et al. \cite{henderson1993experiments} suggested that the the staging behavior could be explained by the change in the value of integer $N$ in Powell's analytical expression. This change could be attributed to the mode switch in the jet. Some other studies by Henderson \cite{henderson1996sound,henderson2002connection} also focused on highlighting the difference between small and large plate impingement. The existence of impinging tones for perfectly expanded jets was reported by Krothapalli et al \cite{krothapalli1999flow}. This was a crucial result since it was believed that the placement of the plate in the shock cell was somewhat important for determining impinging tones. This was also the first study that highlighted the effect of the phase lag term. 

Henderson et al.\cite{henderson2005experimental} pin-pointed the location of the acoustic source as well as the reason for its generation. They provided detailed analysis to prove that the pulsation of the wall jet at a distance of $1.3D$ (where $D$ is the exit diamter of the jet) from the jet centerline produced acoustic waves. The acoustic source was found to be at the same location by Weightman et al \cite{weightman2017explanation}. However, they suggested that the shocklet formation and its subsequent movement generated acoustic waves. Gojon and Bogey \cite{gojon2016investigation} found the source of the acoustic wave at the impingement point.   

To study the upstream traveling feedback acoustic wave in detail, Lepicovsky and Ahuja \cite{lepicovsky1985experimental}had an interesting arrangement. They provided a co-flow velocity with the jet. They observed that the different values of the co-flow velocity did not have any effect on the impinging tones. They hinted at a possible mechanism traveling upstream inside the jet closing the feedback loop. This was later verified by Tam and Ahuja \cite{tam1990theoretical} with their vortex sheet model in which an upstream traveling component called the neutral acoustic mode of the jet was found out. This was extended to supersonic flow by Gojon and Bogey \cite{bogey2017feedback} in which the upstream traveling neutral acoustic mode of the jet was visualized in the density field.

A majority of experimental studies focused on impinging tones and their correlation with Powell's equation, which could also provide some information on any mode switches during impingement. Numerical study of  broadband noise and far-field propagation in impinging jets have not received much attention as compared to free jets. Few numerical studies have focused on these effects with normal as well as oblique impingement. LES of supersonic underexpanded jet impinging  on a large flat plate was reported by Dauptain et al. \cite{dauptain2010large}. They validated their results against existing literature, however, their main aim was to estimate the numerical cost of accurately resolving the essential features in the supersonic jet impingement process using an LES. Numerical studies for oblique impingement were performed by Nonomura et al. \cite{nonomura2011aeroacoustic} and Tsutsumi et al. \cite{tsutsumi2011numerical}. These authors have shown that the sound generation from jet impingement can roughly be divided into three components-acoustic waves generated from the impingement region, Mach wave radiation from the jet shear layer, and Mach wave radiation from the wall jet. LES of perfectly expanded heated and unheated jets impinging normally on a flat plate was performed by Uzun et al \cite{uzun2013simulation}. They also performed a Dynamic Mode Decomposition (DMD) of the data to reveal large coherent axi-symmetric structures at the impingement frequency.
 
The present work aims to study the noise sources and their propagation for the supersonic jet impingement process from an ideally expanded planar nozzle impinging on a flat plate. The evolution of unsteady flow features as well as the noise characteristics are studied when the distance of the plate from the nozzle exit is varied. The present study employs the test cases reported by Gojon and Bogey \cite{gojon2016investigation}. A hybrid methodology \cite{arya2021effect} is employed to study the far-field noise. To the authors' best knowledge, the literature for far-field noise using a hybrid approach in supersonic impinging jets incorporating mean flow effects is very scarce. The significance of the jet oscillations during impingement and the phase lag term in Powell's Equation is also highlighted.

\section{\label{sec:level1}Numerical Method}
All the simulations in the present study are performed using a solver called $rhoEnergyFoam$ \cite{modesti2017low} in the OpenFOAM \cite{jasak2007openfoam} framework. OpenFOAM is an unstructured solver employing the Finite Volume Method. The unsteady compressible Navier-Stokes Equations are represented as- 
  
\begin{equation}
\frac{d}{dt}\int_V\vec{U}dV + \sum_{i = 1}^{3} \int_\Omega (\vec{F}_{i}^{c}-\vec{F}_{i}^{v})nd\Omega = 0
\end{equation}
where
\begin{equation}
\begin{aligned}
\vec{U}=\begin{Bmatrix} \rho \\ \rho u_{i} \\ \rho E\end{Bmatrix},\vec{F}_{i}^{c}=\begin{Bmatrix} \rho u_{i} \\ \rho u_{i}u_{j} + p\delta_{ij} \\ \rho u_{i}H \end{Bmatrix},\\ 
\vec{F}_{i}^{v}=\begin{Bmatrix} 0 \\ \sigma_{ij} \\ \sigma_{ik}u_{k} - q_{i} \end{Bmatrix}
\end{aligned}
\end{equation} represent conservative flux, inviscid flux, and viscous flux, respectively; $u_{i}$ is the velocity component along the $ith$ cartesian coordinate, $\rho$ is the density, E is the total energy, and H is the total enthalpy.

 Advection Upstream Splitting Method (AUSM) \cite{liou1993new} is used for convective flux discretization. An artificial diffusion is added to the solution depending on the Mach number of the flow. For time discretization, the solver employs a low storage, third order, four-stage Runge-Kutta method. This solver has been employed for simulation of many cases covering subsonic as well supersonic flow regimes \cite{modesti2017low, arya2021effect}.

The far-field noise for the present case is calculated using a hybrid approach . In this method, the acoustic sources during runtime are transferred from the fluid flow solver to an acoustic solver which solves the propagation of the acoustic pressure into the far-field. This is achieved with the help of a coupling library. The details of the coupling method can be found out in Arya and De \cite{arya2021effect}. The acoustic solver employed in the current study solves Acoustic Perturbation Equations \cite{ewert2003acoustic}, available in nektar++ \cite{ cantwell2015nektar++}, an open-source higher-order spectral/hp elements framework employing Discontinuous Galerkin Scheme \cite{cockburn2000development}. The APE system is given by 
\begin{equation}
\frac{\partial p'}{\partial t} + \overline c^{2}\nabla \cdot \left(\overline \rho \mathbf{u}' + \overline{\mathbf{u}} \frac{p'}{\overline c^{2}} \right) = \overline{c}^{2}q_{c}
\end{equation}

\begin{equation}
\frac{\partial \mathbf{u'}}{\partial t} + \nabla \left(\overline{\mathbf{u}} \cdot \mathbf{u}' \right) + \nabla \left( \frac{p'}{\overline \rho} \right)  = \mathbf {q_{m}}
\end{equation}
where the sources are given by-
\begin{equation}
q_{c} = -\nabla \cdot \left(\rho' \mathbf{u}' \right)' + \frac{\overline \rho}{C_{p}} \frac{Ds'}{Dt}
\end{equation}

\begin{equation}
\mathbf{q_{m}} = -\left(\mathbf{\omega} \times \mathbf{u} \right)' + T'\nabla \overline s - s'\nabla \overline T - \left(\nabla \frac{u'^{2}}{2} \right)' + \left(\frac{\nabla \cdot \overline \tau}{\rho} \right)'
\end{equation}

The primed variables are the fluctuating variables, whereas an overbar represents the time-averaged quantities. For the present case, the fluctuating Lamb Vector $(\omega \times u)'$ and the spatio-temporal changes in fluctuating entropy are used as source terms for all the simulations.

\section{Computational Details}

\begin{figure}
	\includegraphics[scale=0.8]{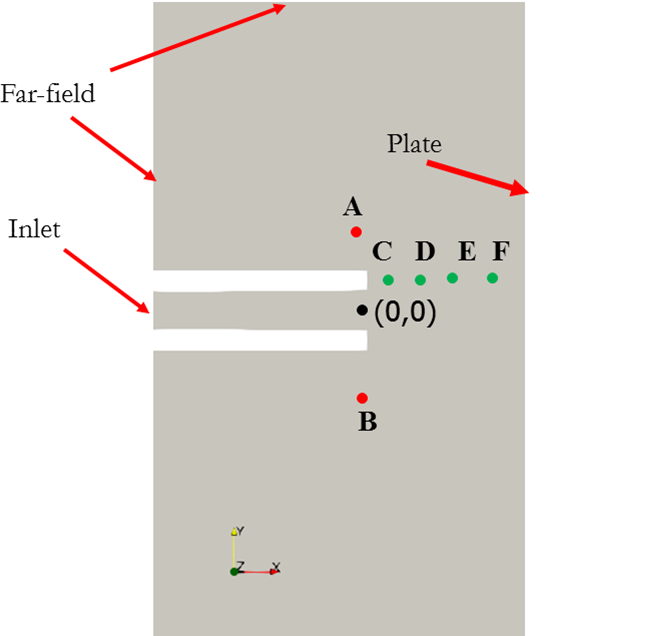}
	\centering
	\caption{\label{fig:4.2}Schematics of Geometry and Computational details}	
\end{figure}
The representative geometry for all the cases is presented in Fig. \ref{fig:4.2}. The exit Mach number is $1.28$ and is fixed for all the cases. The diameter of the nozzle exit (D) is $12.7$ mm. The lip thickness is $D/2$ and the Reynolds number of the incoming flow is $5e4$. Four cases with different nozzle to plate distance and perpendicular impingement, and one case with angled impingement ($75deg$) are simulated for the present study. The case number is reported according to the distance of the plate from the nozzle. These details are provided in Tab. \ref{table:4.1}. The pressure field is probed at points A and B for correlation analysis. Velocity values are probed at points C, D, E, F for calculating the convective velocity of the large-scale structures.

The present case employs a hybrid method for far-field noise calculation. It employs different grids and different time steps for both solvers. The grid generation process and the coupling strategy are explained in detail in Arya and De \cite{arya2021effect}. These strategies determine the accurate resolution of all the dominant frequencies in the flow. All the grids in the present case have been generated following these strategies. The far-field location and the number of grid points for each case are presented in Tab. \ref{table:4.2}.

The inlet turbulence is generated through a digital filter  \cite{klein2003digital} based method. A turbulence level of $6\%$ is provided at the inlet. WALE model is used for the resolution of SGS stresses. The inlet turbulence generation method is different in the reference study. Nevertheless, it is observed in the later section that a good match is obtained for mean turbulent fluctuations. Wave-transmissive boundary conditions are applied at the far-field boundary of the fluid \cite{soni2017investigation}. The far-field boundary of the acoustic domain uses a Reimann-invariant characteristic boundary condition which suppresses any incoming characteristics into the domain. The walls in the fluid domain are no-slip, whereas in the acoustic domain, they are prescribed with a fully reflecting boundary condition. Periodic conditions are imposed at the spanwise planes. With this set up, the present case resembles a rectangular jet with a high aspect ratio.  

\begin{table}
	\caption{Cases considered for the current study}
	\begin{tabularx}{\textwidth} {|X|X|X|}
		\hline
		\textbf{Case}&\textbf{Distance}&\textbf{Angle (degree)}  \\
		\hline
		x/D-3pt94 & 3.94D & 0 \\
		\hline 
		x/D-5pt5 & 5.5D & 0  \\
		\hline 
		x/D-8pt27 & 8.27D & 0  \\
		\hline 
		x/D-9pt1 & 9.1D & 0  \\
		\hline 
		x/D-5pt5-75deg & 5.5D & 75 \\
		\hline 
	\end{tabularx}
	\label{table:4.1}
\end{table}

\begin{table}
	\caption{Grid Details for different cases}
	\begin{tabularx}{\linewidth}{|X|X|X|X|X| }
		\hline
		\textbf{Case}&\textbf{Fluid boundary}&\textbf{Acoustic boundary}&\textbf{Fluid grid}&\textbf{Acoustic grid}  \\
		\hline
		x/D-3pt94 & 210D & 200D& 3.0e+6&7.5e+5 \\
		\hline 
		x/D-5pt5 & 210D & 200D& 3.5e+6&8.0e+5  \\
		\hline 
		x/D-8pt27 & 210D & 200D& 4.0e+6& 9.0e+5  \\
		\hline 
		x/D-9pt1 & 210D & 200D& 4.0e+6&9.0e+5 \\
		\hline 
		x/D-5pt5-75deg & 210D & 200D& 3.5e+6&8.0e+5 \\
		\hline 
	\end{tabularx}
	\label{table:4.2}
\end{table}

\section{Results and Discussion}

\begin{figure}
	\includegraphics[scale=0.45]{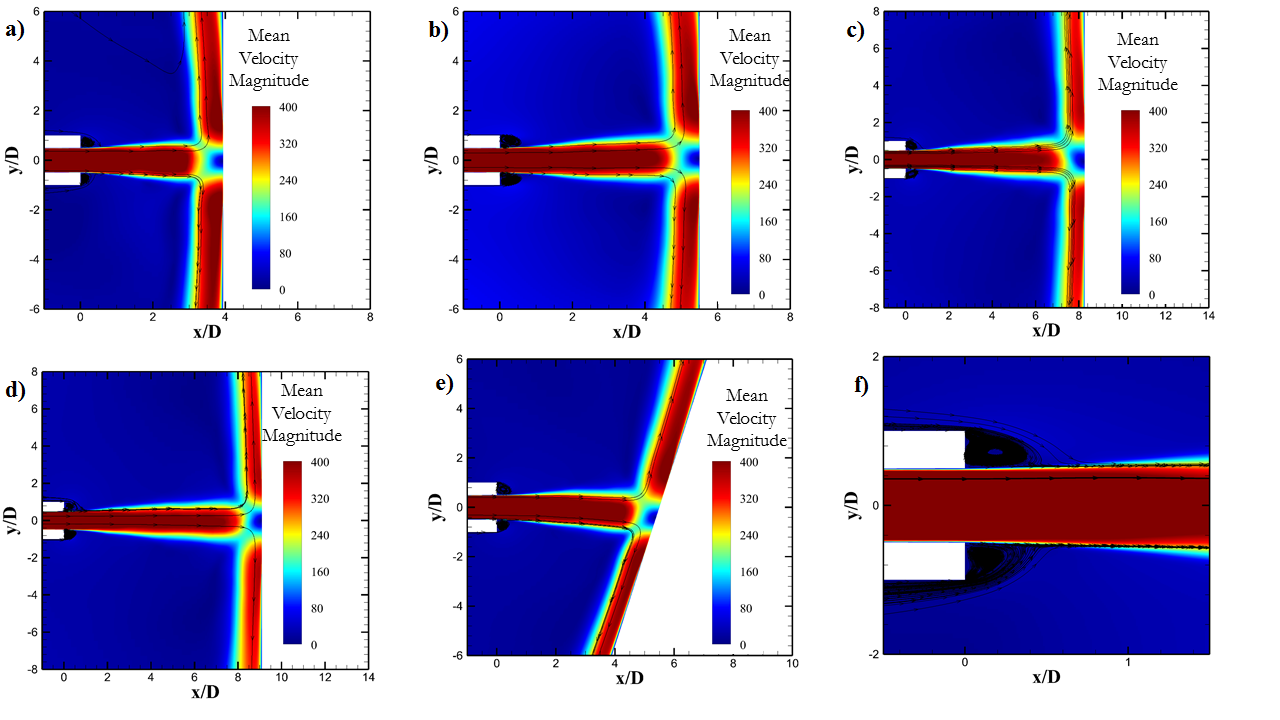}
	\centering
	\caption{\label{fig:4.3} Mean Velocity Magnitude(m/s) a) x/D-3pt94, b) x/D-5pt5, c) x/D-8pt27, d) x/D-9pt1, e) x/D-5pt5-75deg, f) x/D-3pt94-close up view}
\end{figure}

The mean velocity magnitude contours for all the cases are presented in Fig.  \ref{fig:4.3}. The essential flow features are similar for all the cases. Though the jet is ideally expanded, due to the entrainment of the fluid upstream of the nozzle, the pressure at the exit of the nozzle slightly decreases causing a slight mismatch from the ideally expanded conditions. Nevertheless, this deviation  is not significant and no shock cells are developed. Consequently, the jet, in this case, may be treated as an ideally expanded jet. The entrainment of the fluid upstream of the jet also leads to the development of recirculation regions in the lip region. This is highlighted in Fig. \ref{fig:4.3}(f), which shows a close-up view near the nozzle exit. A stand-off shock is developed near the plate when the supersonic jet impinges on the plate. After impinging on the plate, the flow moves nearly parallel to the wall, visible from the streamlines drawn over each figure. The flow profile resembles a jet issuing near the impingement point and along the plate. Thus, it is called a wall jet. 

\begin{figure}
	\includegraphics[scale=0.2]{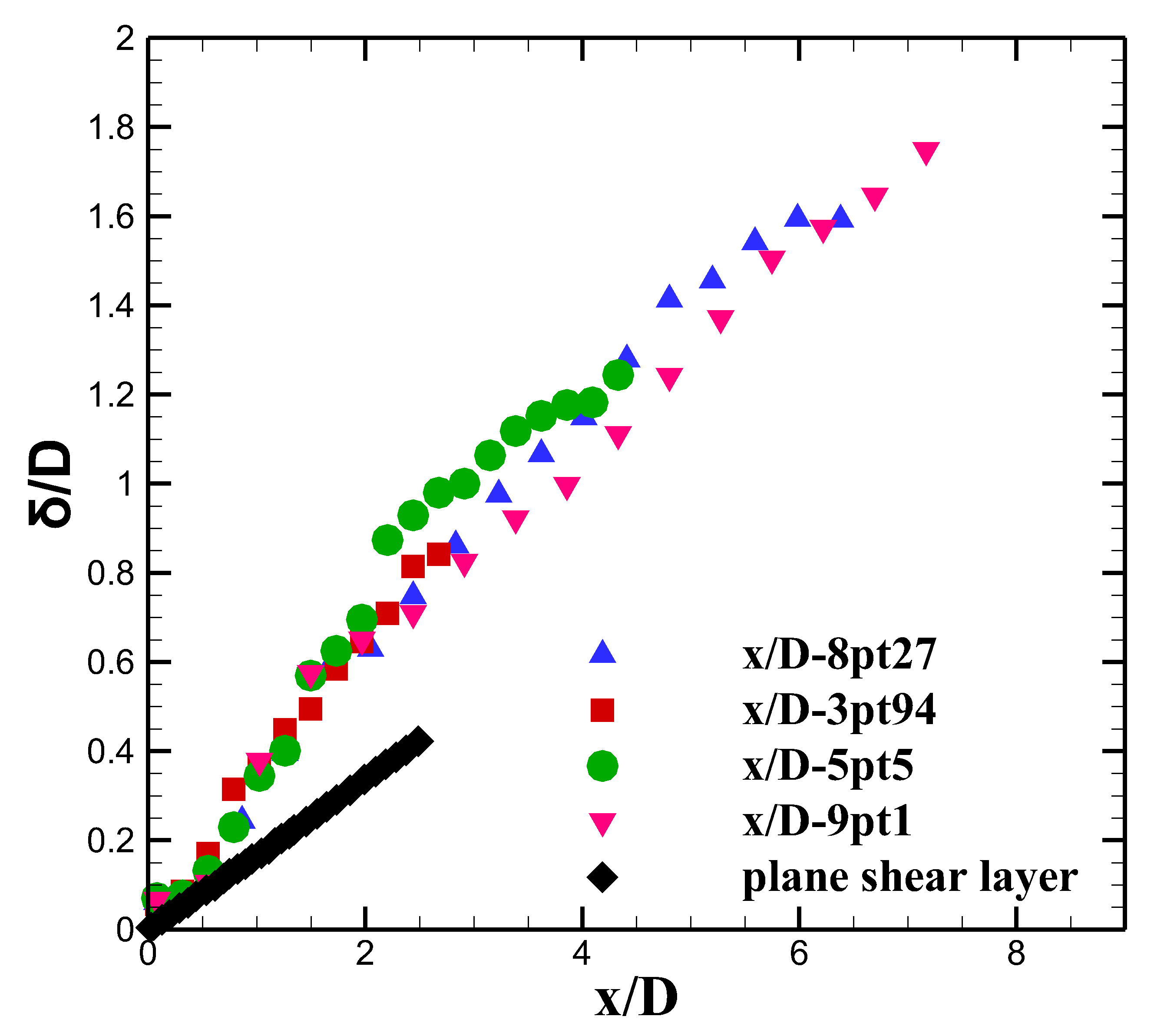}
	\centering
	\caption{\label{fig:4.4} a) Mean Shear-layer thickness for different cases  }
\end{figure}
\begin{figure}
	\includegraphics[scale=0.6]{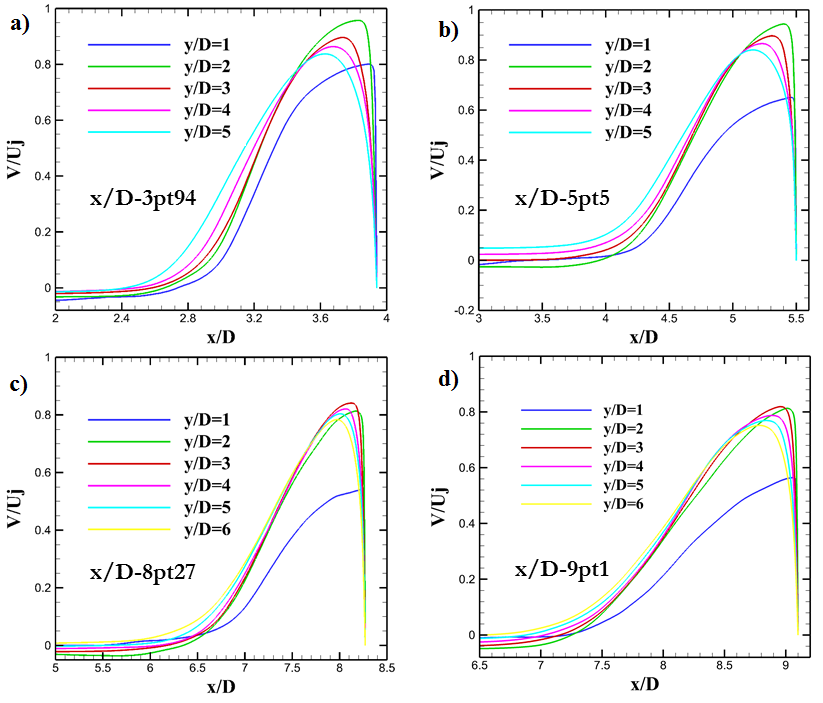}
	\centering
	\caption{\label{fig:4.5} Wall jet profiles for different cases: a)x/D-3pt94, b) x/D-5pt5, c) x/D-8pt27 d) x/D-9pt1}
\end{figure}

The shear layer thickness for different cases is presented in Fig. \ref{fig:4.4}. The shear layer thickness is defined in this case as $R_{0.05}-R_{0.95}$, where $R_{0.05}$ is the point along the $y-axis$ where the axial velocity is $5\%$ of the local jet centerline axial velocity and $R_{0.95}$ is the point along the $y-axis$ where the axial velocity is $95\%$ of the local jet centerline axial velocity. It can be observed that near the separation region at the nozzle exit, the shear layer thickness is very less and is comparable to a plane mixing layer \cite{castro1973highly}. As we move downstream of the nozzle, the shear layer thickness increases. However, the growth rate is larger than that of a plane mixing layer. This is attributed to the turbulence and 3D nature of the flow. The growth rate for all the cases is more or less linear.

One of the striking features of a jet impinging on a flat surface is the formation of a wall jet parallel to the plate. The profiles of wall jet along $x-direction$ at different $y$ locations for different cases are presented in Fig. \ref{fig:4.5}. The velocity along $y-axis$ is normalized by the jet centerline velocity, $U_j$. For smaller plate distances (Fig. \ref{fig:4.5}(a) Fig. \ref{fig:4.5}(b)), we see that the peak velocity along the $y$ direction is more than $0.9U_j$ whereas, for larger plate distances(Fig. \ref{fig:4.5}(c) Fig. \ref{fig:4.5}(d)),  the peak velocity along the $y$ direction is nearly $0.8U_j$. The peak velocity decreases rapidly for the smaller plate distances as we move along the $y-axis$. Also, the peak occurs at lower $y/D$ values for smaller plate distances. However, we can see that the peak velocity value occurs at a very short distance from the plate  for all the cases.

\begin{figure}
	\centering
	\includegraphics[scale=0.5]{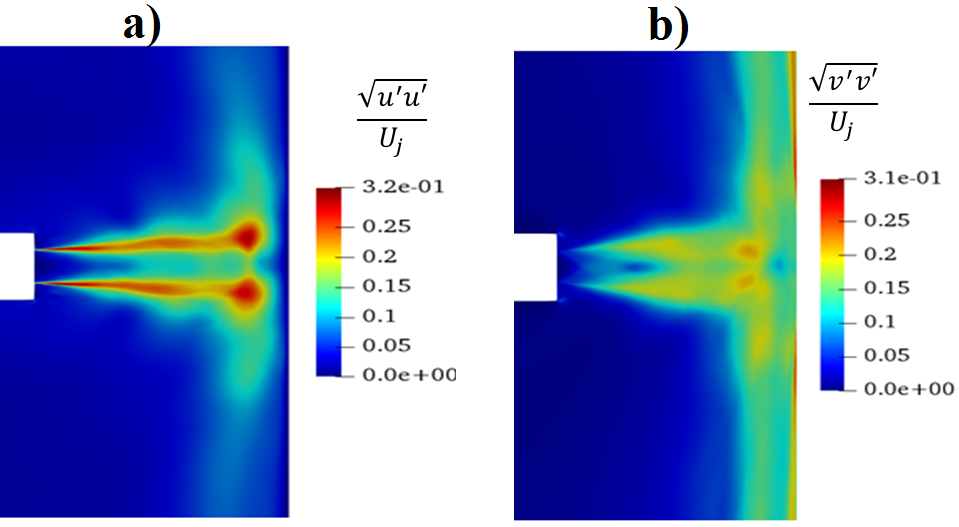}
	\centering
	\caption{\label{fig:4.6} Velocity Perturbations a) along $x-axis$, b) along $y-axis$}
\end{figure}

\begin{figure}
	\centering
	\includegraphics[scale=0.5]{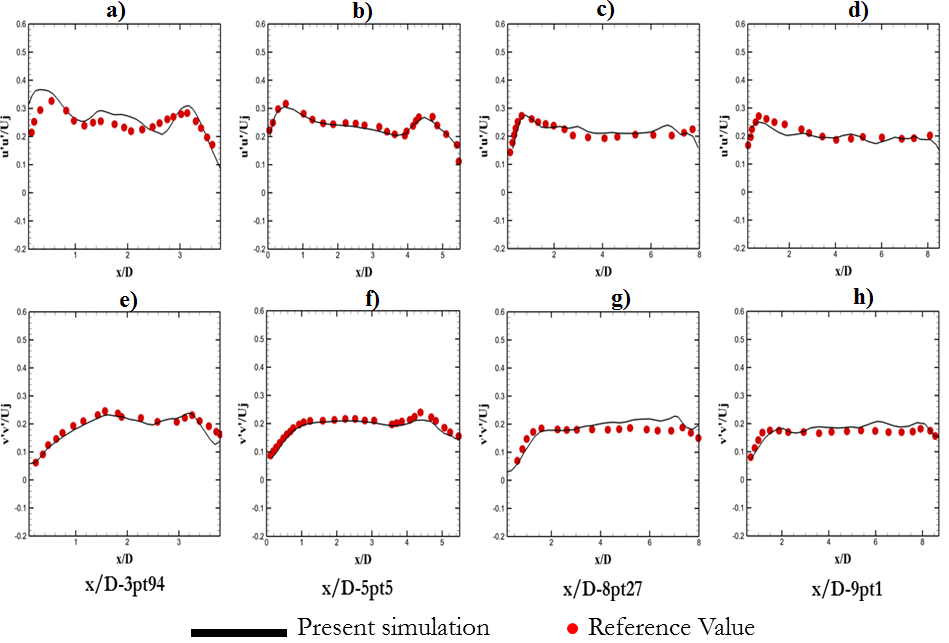}
	\caption{\label{fig:4.7} Time averaged Axial- a) x/D-3pt94, b) x/D-5pt5, c) x/D-8pt27, d) x/D-9pt1 and Normal- e) x/D-3pt94, f) x/D-5pt5, g) x/D-8pt27, h) x/D-9pt1 velocity fluctuations}
\end{figure}

The mean velocity fluctuations along axial (Fig. \ref{fig:4.6}(a)) and normal (Fig. \ref{fig:4.6}(b)) direction for one of the cases (x/D-5pt5) highlight different features in the flow. High axial fluctuations are observed near the nozzle lip, along the shear layer and near the wall jet. A Slightly different pattern is observed when we consider the normal fluctuations. The values near the jet are small which progressively increase as we move downstream . The highest values of normal fluctuations are found near the wall jet. This distribution is also quantitatively assessed while comparing against the reference study \cite{gojon2016investigation} and is presented in Fig. \ref{fig:4.7}. It can be seen that the values predicted in the current simulation are fairly close to the that reported in the reference study. However, few deviations are observed resulting from the imperfect representation of the status of turbulence in the current study compared to the reference. Nevertheless, the locations of the peaks are predicted well in the present simulation. The peak in axial fluctuations (from Fig. \ref{fig:4.7}(a)-(d)) at the nozzle exit is due to the excitation of the shear layer at the lip by the feedback acoustic waves. However, the normal fluctuations (from Fig. \ref{fig:4.7}(e)-(h)) are lower at this point which suggests that the excitation is predominantly along the axial direction. Both axial and normal fluctuations are higher near the starting of the wall jet region. This is probably due to high shear developed from the interaction of the jet shear layer with the wall and with the developing wall jet.


\begin{figure}
	\includegraphics[scale=0.5]{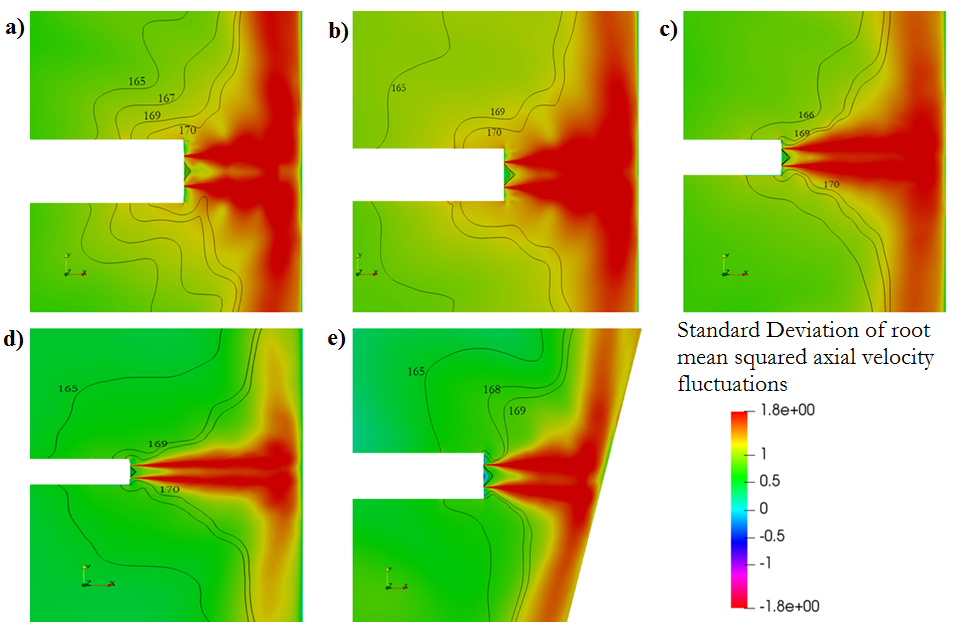}
	\centering
	\caption{\label{fig:4.8} Standard deviation of rms velocity fluctuations-  shown in color, and OASPL(dB)-shown as line contours : a) x/D-3pt94, b) x/D-5pt5, c) x/D-8pt27, d) x/D-9pt1, e) x/D-5pt5-75deg}	
\end{figure}

The feedback process developed in the impinging jet scenario gives rise to an aeroacoustic resonance. This can also be understood from a simple observation that the downstream propagating structures constitute a downstream traveling wave. When it encounters an upstream traveling acoustic wave, an interference takes place. This interference can create resonance conditions between the plate and the nozzle. The aeroacoustic resonance had previously been visualized qualitatively in density gradient field \cite{westley1975near}. However, in the present case, we visualize it with the help of the standard deviation of root mean squared axial velocity fluctuations. Since the spread of this quantity over the domain is huge, we visualize it on a logarithm scale. This is presented in Fig. \ref{fig:4.8}. The contour lines of the Overall Sound Pressure Level (OASPL) are also presented. The standing wave pattern is visible for smaller plate distances (Fig. \ref{fig:4.8}(a) and Fig. \ref{fig:4.8}(b)) in the form of lobes across the axis. This pattern is not clearly visible for larger plate distances (Fig. \ref{fig:4.8}(c) and Fig. \ref{fig:4.8}(d)). This means that there is a strong hydrodynamic-acoustic coupling for smaller plate distances. This strong coupling results in periodic structures found in the system. Interestingly, for angled impingement (Fig. \ref{fig:4.8}(e)), a slight breakdown of the standing wave pattern is observed just as in the cases with a larger plate distance. This means that a change in angle of the plate disrupts the hydrodynamic-acoustic coupling for the impinging jets, probably caused due to the change in the pattern of the feedback acoustic waves generated. It can be observed from the OASPL contours presented in Fig. \ref{fig:4.8}(e) that for angled impingement, the resulting sound-field has an asymmetric pattern.  The results for normal injection clearly show a symmetric sound field. However, it is clear for all the cases that multiple sound sources are located. The first acoustic component is seen at an angle less than $30 degrees$. This acoustic component is responsible for the feedback acoustic waves. Upstream of the nozzle exit, the propagation direction of this component makes a small angle from the nozzle centerline. This is probably due to the effect of the reflection of waves from the nozzle walls. The other acoustic component is located at an angle greater than $45 degrees$ from the nozzle axis. The source of this component seems to be different from the source of the acoustic component discussed previously. While the source for the former acoustic component lies in the impingement region, the source for the latter component is at the wall jet region. The OASPL here sheds some light on different acoustic sources located at the impingement region and at the wall jet region. The picture would be clearer if we observe the unsteady flow field near the nozzle.

\subsection{Unsteady Flow Features and Noise Field}

\subsubsection{Near-Field Characteristics}

\begin{figure}
	\begin{subfigure}{1\textwidth}
		\includegraphics[scale=0.5]{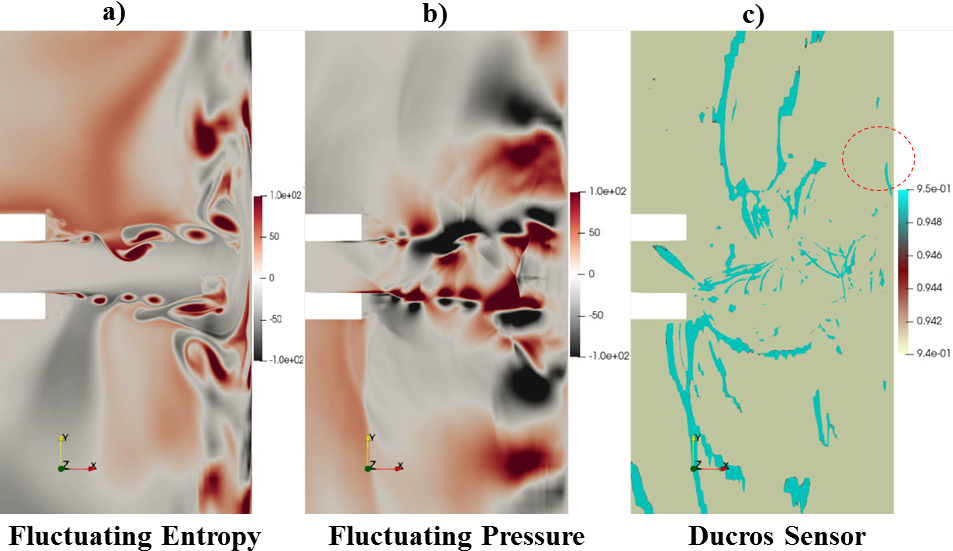}  
		\centering
	\end{subfigure}
	
	\begin{subfigure}{1\textwidth}
		\includegraphics[scale=0.5]{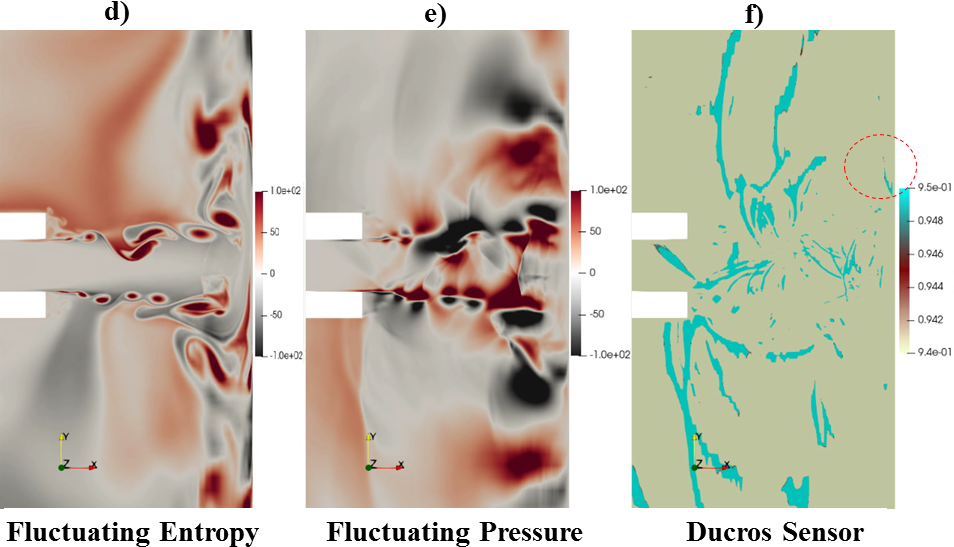}
		\centering  
	\end{subfigure}
	
	\begin{subfigure}{1\textwidth}
		\includegraphics[scale=0.5]{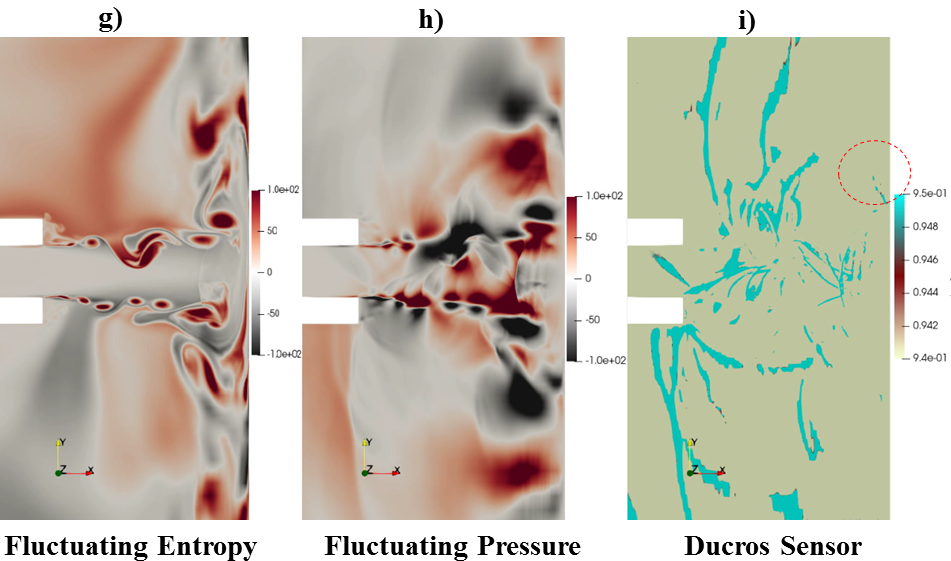}  
		\centering
	\end{subfigure}
	\caption{Unsteady Flow features}
	\label{fig:4.9}
\end{figure}

The supersonic jet impinging on a flat surface is a highly unsteady phenomenon. A clear understanding of the flow process and the sound generation can be explained through the application of innovative techniques on the unsteady flow field. For instance the downstream convection of shear layer vortices can most aptly be represented by analyzing the evolution of vorticity in time. However, it is found in the present study that the fluctuating entropy contours highlight the evolution of these vortices even along the wall jet region, where vorticity contours do not reveal much information. The fluctuating entropy is used as one of the source terms for far-field noise calculation discussed in a later section. The change in entropy at every computational point is calculated using the state variables with value in the previous iteration as the base state. We can see (from Fig. \ref{fig:4.9}(a), Fig. \ref{fig:4.9}(d), Fig. \ref{fig:4.9}(g)) the evolution of shear layer vortices as they convect downstream. The size of the vortices increase while they are convected from the nozzle exit to the plate. As they move along the wall jet, a pulsation of the wall jet occurs due to the interaction of the vortices with the plate. This pulsation of the wall jet had previously been reported by Henderson et al. \cite{henderson2005experimental} as the source of acoustic waves. The location of the pulsation is found to be nearly at $y = 1.38D$ which is fairly close to the value reported in the experiment. 

The near field pressure fluctuation contours which are composed of hydrodynamic as well as acoustic components are presented in Fig. \ref{fig:4.9}(b), Fig. \ref{fig:4.9}(e), Fig. \ref{fig:4.9}(h). The downstream convection of vortices along the shear layer is visible. The obstruction created in the supersonic flow by large-scale structures generates variety of waves in the system. Therefore, the acoustic waves in these contours are bit unclear. However, the upstream propagating feedback waves near the nozzle exit are visible.

The discontinuities in the flow are revealed using a shock sensor called Ducros sensor \cite{ducros1999large}. This sensor is based on dilatation and is used to detect the regions where shocks are located in the flow. The dilatation field also represents the compressible fluctuations in the flow, related to the temporal derivative of pressure. Thus, the shock sensor can also provide a glimpse of the sound field in the flow. The Ducros sensor is presented in Fig. \ref{fig:4.9}(c), Fig. \ref{fig:4.9}(f), Fig. \ref{fig:4.9}(i). It reveals some weak shocks developed near the nozzle and also the feedback waves that emanate from the plate. However, the main motive to employ Ducros sensor in this case is to particularly study the small shocklet that is highlighted in the red circle. The images are taken for one cycle of the feedback loop. The shocklet is seen to rotate in the upstream direction. After its rotation, it is no longer captured in the Ducros sensor contours until a new shocklet appears and the same process repeats. This phenomenon is also observed at the same negative $y$-location. This process has been reported by Weightman et al. \cite{weightman2017explanation}. The authors observed the generation of acoustic waves after the movement of the shocklet. In the present case, however, the generation of the acoustic wave cannot be determined due to the inadequate resolution in representing the shocklet inception and its movement.

\begin{figure}
	\begin{subfigure}{1\textwidth}
		\includegraphics[scale=0.6]{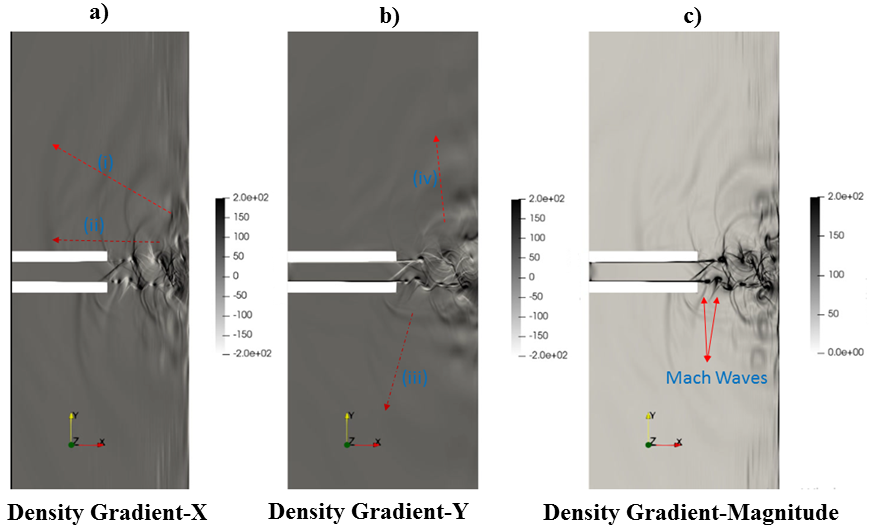}  
	\end{subfigure}
	
	\begin{subfigure}{1\textwidth}
		\includegraphics[scale=0.58]{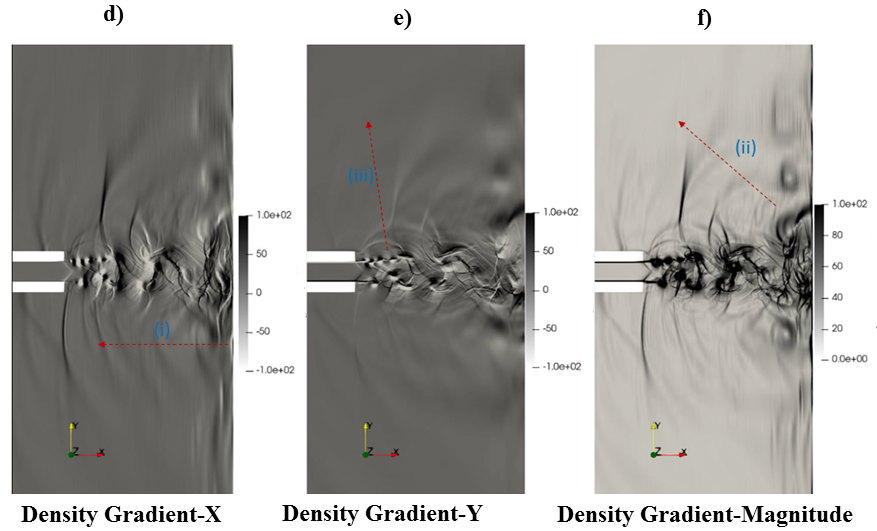}  
	\end{subfigure}
	
	\begin{subfigure}{1\textwidth}
		\includegraphics[scale=0.58]{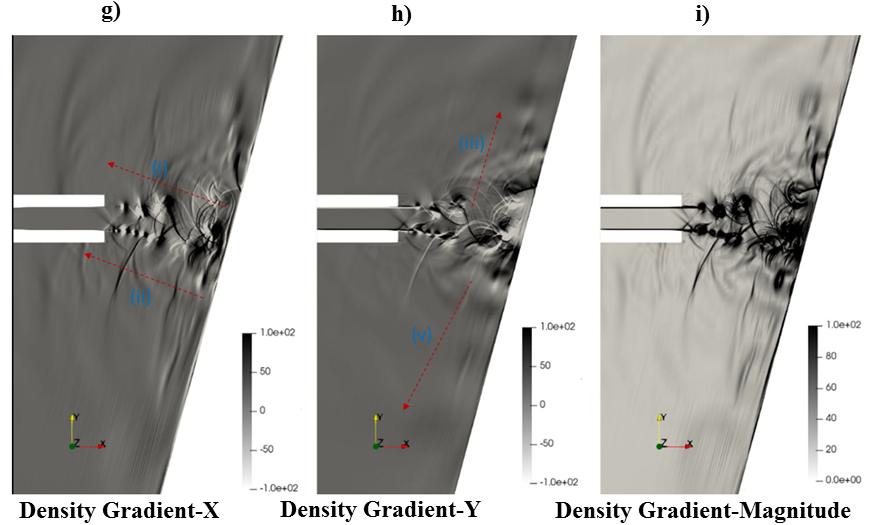}  
	\end{subfigure}
	
	\caption{Density Gradient contours: arrows (i),(ii)- feedback waves, (iii),(iv) broadband acoustic waves, (v) waves due to interaction at wall jet}
	\label{fig:4.10}
\end{figure}

We also present density-gradient contours for further analysis of near field flow structures and associated noise. These contours are presented in Fig. \ref{fig:4.10}. The contours reveal small shock structures near the nozzle exit. As already discussed this happens by reduced pressure at the exit of the nozzle developed due to entrainment of upstream fluid. However, the pressure does not reduces significantly and there are no shock cells near the exit which removes the possibility of jet screech \cite{raman1998advances,powell1953noise,berland2007numerical,rao2020screech} for the present case. The near-field of a supersonic impinging jet is dominated by a large number of waves. One of the striking features of supersonic jets is Mach wave radiation \cite{mclaughlin1975experiments,tam1992relationship, mitchell1995direct}. The large-scale structures in the supersonic shear layer move at very high speeds described by their convective velocities. Mach waves are generated when these structures move supersonically with respect to the ambient air. Since a multitude of large scale structures exist in a typical shear layer, a supersonic jet is filled with Mach wave radiation. This leads to a high level of broadband noise in supersonic jets \cite{papamoschou1997mach,tsutsumi2011numerical}. For the present case, the flow Mach number is not sufficiently high due to which the convection velocities are not supersonic at all points downstream of the jet exit. Mach waves can be seen emanating from the shear layer at the nozzle exit in Fig. \ref{fig:4.10}(f) (case - x/D-3pt94). However, acoustic waves due to Mach radiation are not significant in the present case. This is due to the scarcity of Mach waves for the present case as well as due to masking of the acoustic radiation due to Mach waves from other waves present in the system. 

Various features are explained by observing three different cases corresponding to smaller plate distance (x/D-3pt94 - Fig. \ref{fig:4.10}(a), Fig. \ref{fig:4.10}(b), Fig. \ref{fig:4.10}(c)), larger plate distance (x/D-8pt27 - Fig. \ref{fig:4.10}(d), Fig. \ref{fig:4.10}(e), Fig. \ref{fig:4.10}(f)), and angled impingement (x/D-5pt5-75deg - Fig. \ref{fig:4.10}(g), Fig. \ref{fig:4.10}(h), Fig. \ref{fig:4.10}(i)). It is observed that the feedback waves (arrows marked (i) and (ii)) for smaller as well as larger plate distances, are focused particularly along two directions. The first one is anti-parallel to the jet axis, and the second one makes an acute angle with the jet axis. This observation suggests that the source of these waves lie respectively in the impingement point and at nearly (y/D~2.5) from the jet centerline. For x/D-5pt5-75-deg the observation suggests that the high intensity feedback waves are mostly focused perpendicular to the plate. Also, the radiation in the negative $y-axis$ is dominated by broadband sound (marked with arrows (iii), (iv)). For the smaller and larger plate impingement, the broadband sound is emanated equally in either directions. Another component that seems to originate at the wall jet (marked with arrow (v)) can be observed in Fig. \ref{fig:4.10}(h)     

\begin{figure}
	\includegraphics[scale=0.45]{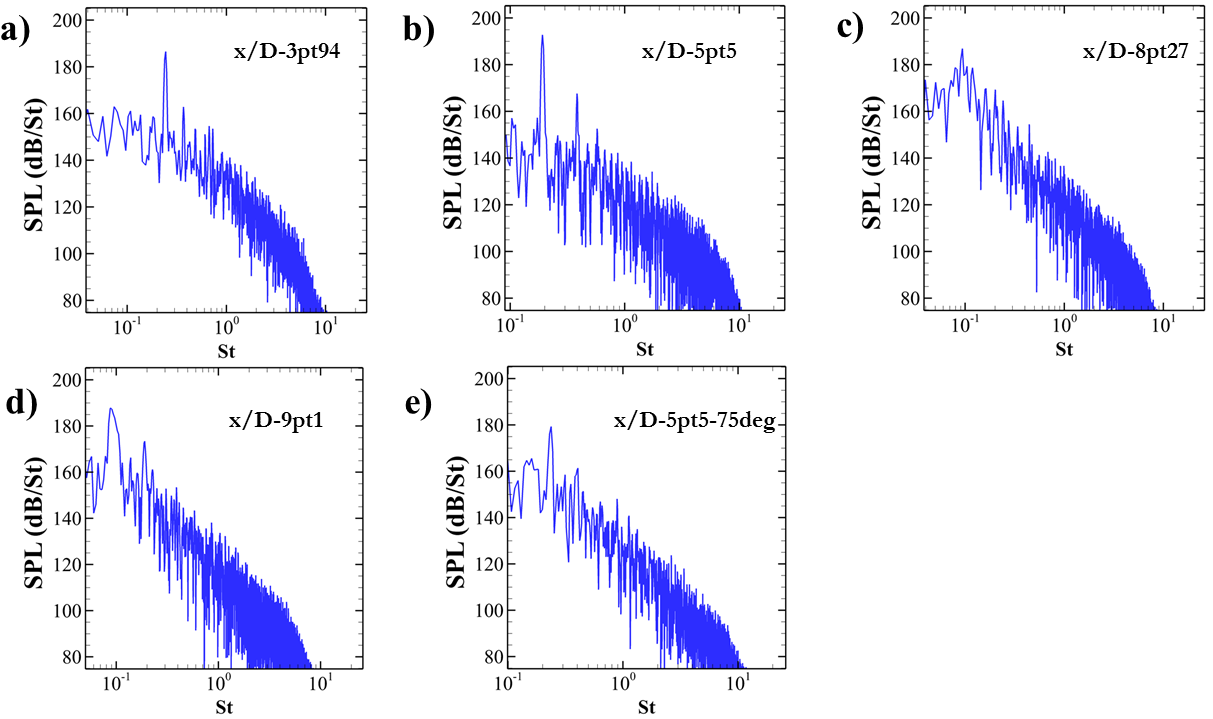}
	\centering
	\caption{\label{fig:4.11} FFT spectrum at $x=0,y=1.5D$ : a) x/D-3pt94, b) x/D-5pt5, c) x/D-8pt27, d) x/D-9pt1, e) x/D-5pt5-75deg}	
\end{figure}

The data presented so far qualitatively gives us a picture of the flow and acoustic fields. The tonal noise generated due to the impingement process is presented now in Fig. \ref{fig:4.11}. These measurements are made at $x=0,y=1.5D$. The existence of multiple  peaks is observed for each case. It should be noted that these tones are produced solely due to the jet impingement process because of the absence of any shock cell structures. The Strouhal number corresponding to the peaks predicted in the present case are highlighted in Tab. \ref{table:4.3}. The Strouhal number which is not captured in the present analysis but obtained in the reference study is left as a blank. The corresponding SPL values (dB/St) for the highest peaks obtained for each case are also shown in parentheses. The FFT spectrum presented, captures the major peaks in all the cases with reasonable accuracy. The value of the Strouhal number for each case is fairly close to that reported in the reference study. We can see that the loudest sound is obtained for x/D-5pt5. The difference between the peak value for this case and the reference study is $5dB$. Other higher harmonics are also observed for this case. As discussed earlier, the distance between the plate and the nozzle exit for the case x/D-5pt5 corresponds to the strong resonance condition responsible for generating maximum sound amplitude. The lowest peak is observed for the case x/D-5pt5-75deg. As noted earlier, the lower half of the domain for the x/D-5pt5-75deg, is dominated by broadband noise. This suggests that there is some disruption of the feedback loop. The change in the direction of feedback acoustic waves was noticed earlier. This results in the different excitation of the shear layer at the upper and the lower lips which is probably responsible for the decrease in tonal noise.

\begin{figure}
	\includegraphics[scale=0.45]{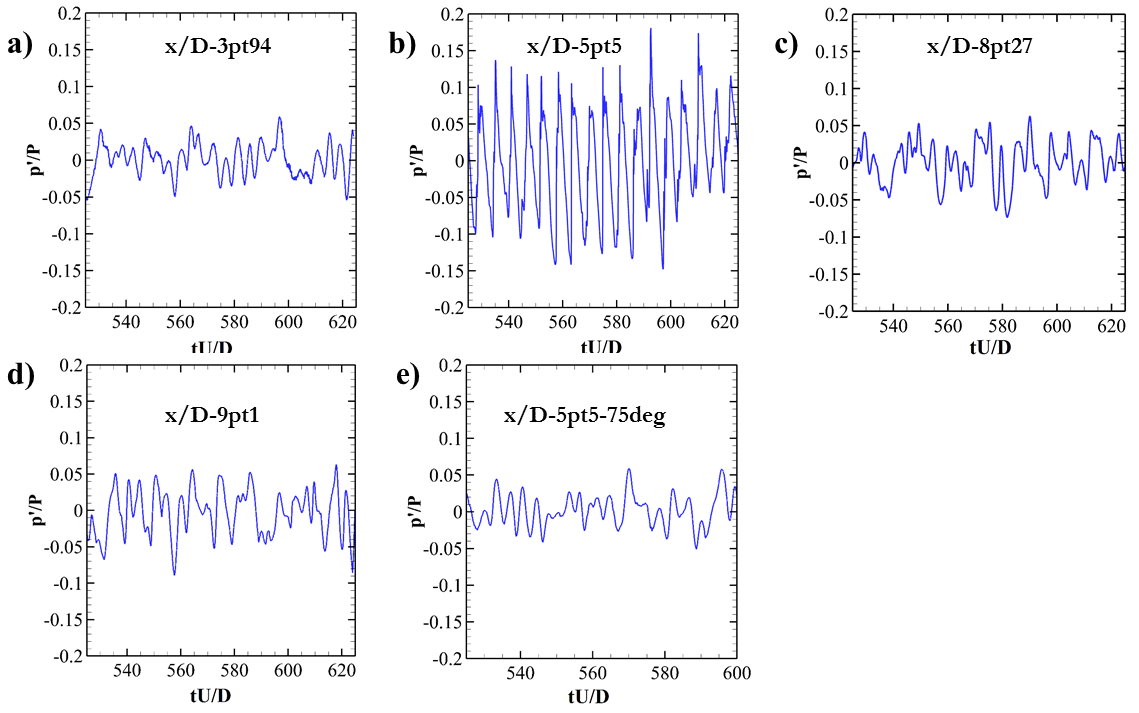}
	\centering
	\caption{\label{fig:4.12} Pressure fluctuations at $x=0,y=8.5D$ : a) x/D-3pt94, b) x/D-5pt5, c) x/D-8pt27, d) x/D-9pt1, e) x/D-5pt5-75deg}	
\end{figure}

\begin{figure}
	\includegraphics[scale=0.5]{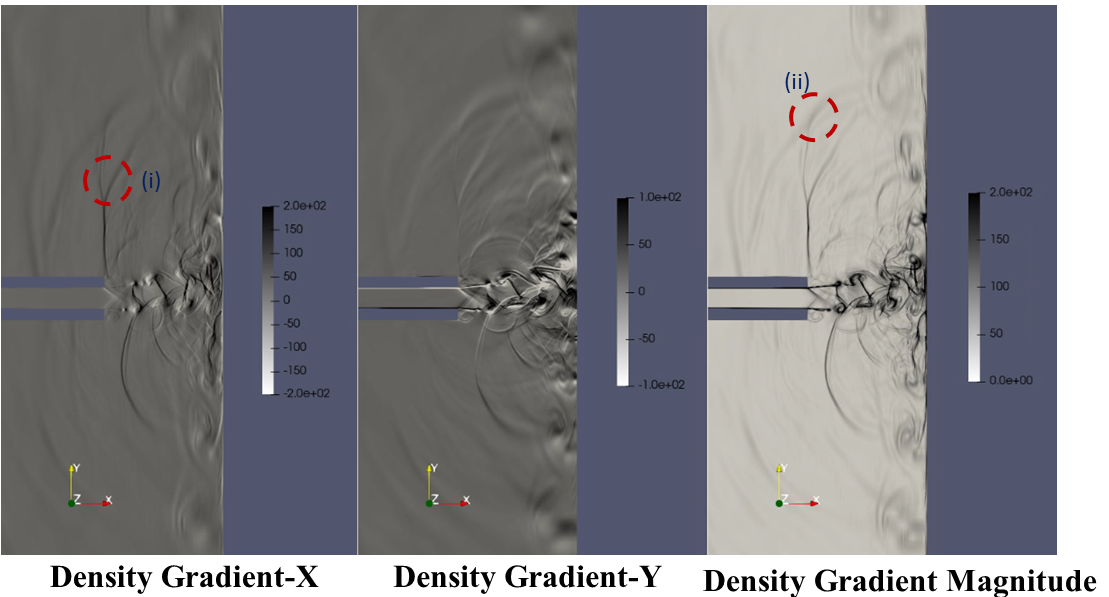}
	\centering
	\caption{\label{fig:4.13} Coalescence of feedback waves for x/D-5pt5}	
\end{figure}

The unsteady pressure field is monitored at $x=0, y=8.5D$ to study any non-linearity in propagation. The pressure field normalized by the ambient pressure is presented in Fig. \ref{fig:4.12}. It is observed that high-pressure fluctuations are obtained for x/D-5pt5. The characteristic shape of these fluctuations suggests the presence of crackle-generated noise \cite{krothapalli2000crackle,williams1975crackle,nichols2013crackle}. The skewness and Kurtosis of the pressure signal are also presented in Tab. \ref{table:4.4}. Skewness represents the deviation of a signal from the normal distribution. It had been suggested by Ffowcs Williams \cite{williams1975crackle} that the pressure signal for crackle associated noise usually displays skewness values greater than 0.4. The skewness value for the pressure signal obtained for x/D-5pt5 is very high. Also, the Kurtosis value for this case is greater than 3. High Kurtosis values are representative of signals displaying higher peak values. Thus, the pressure values at this point suggest the presence of crackle noise. However, crackle noise results from micro explosions generated when a cold stream of air suddenly expands on coming in contact with a hot flow. So the observation of crackle noise is doubtful for the current low supersonic, unheated case. To find out more, density gradient contours are presented in Fig. \ref{fig:4.13}, obtained when a rise in a peak occurs (Fig. \ref{fig:4.12}(b)). The characteristic crackle waves are not seen in these contours. Instead, the coalescence of feedback waves is observed near the sampled point marked with circles (i) and (ii). Although, the coalescence of feedback waves for other cases is also observed, the intensity of these waves is highest for x/D-5pt5, as clear from the previous observations. The merging of these feedback waves can generate a pressure signal similar to one obtained in Fig. \ref{fig:4.12}.

\begin{table}
	\caption{Frequencies obtained for each case}
	\begin{tabularx}{\linewidth}{|X|X|X|X|X|X|}
		\hline
		Case&St1&St2&St3&St4&St5 \\
		\hline
		x/D-3pt94 & ---- & \textbf{0.2443} (187) & 0.3693 & 0.4886 & ----  \\
		\hline 
		x/D-5pt5 & 0.1033 & \textbf{0.1914}(193) & 0.3828 & 0.5780 & ---- \\
		\hline 
		x/D-8pt27 & \textbf{0.0937}(186) & ---- & 0.1611 & 0.2017 & 0.2512 \\
		\hline 
		x/D-9pt1 & \textbf{0.0870}(187) & 0.1386 & 0.1898 & 0.2260 & ---- \\
		\hline 
		x/D-5pt5-75deg & 0.1670 & \textbf{0.2373}(179) & ---- & ---- & ---- \\
		\hline 
	\end{tabularx}
	\label{table:4.3}
\end{table}

\begin{figure}
	\includegraphics[scale=0.45]{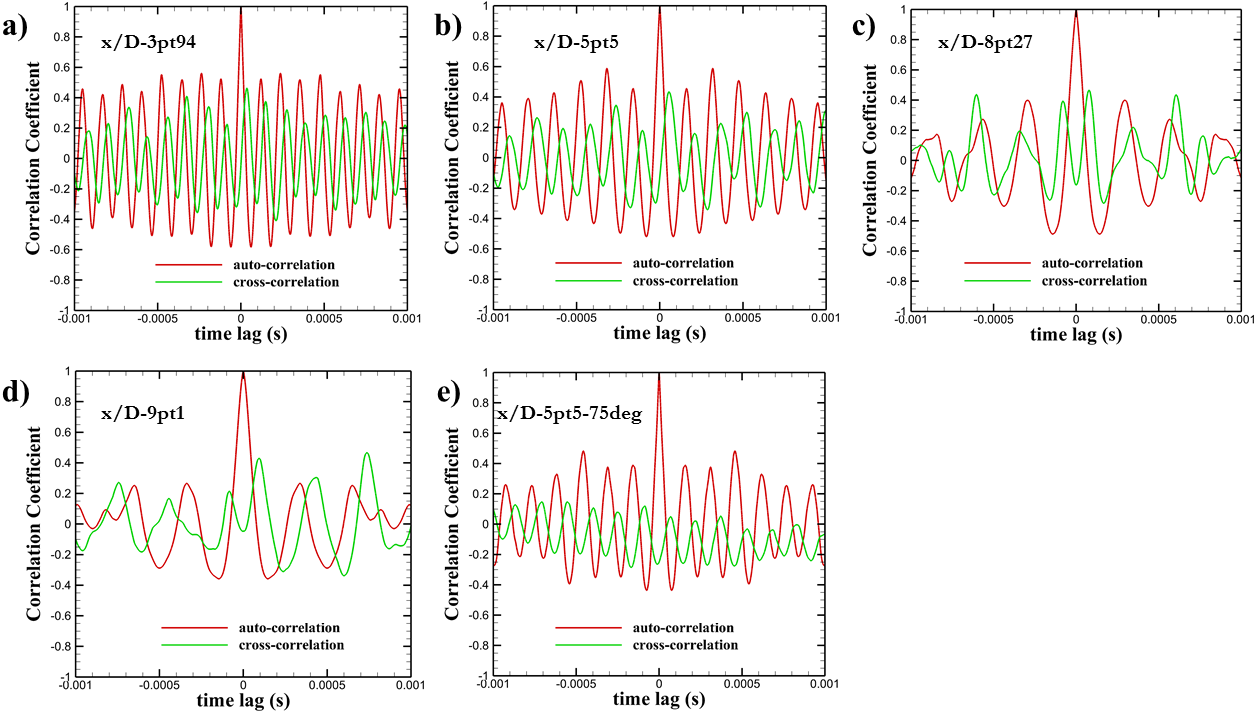}
	\centering
	\caption{\label{fig:4.14}Correlation analysis for different cases:a) x/D-3pt94, b) x/D-5pt5, c) x/D-8pt27, d) x/D-9pt1, e) x/D-5pt5-75deg }	
\end{figure}

We further study the tone generation process with the help of correlation analysis. The pressure values probed at point A, and point B are used for the calculation of cross-correlation for the present case. The correlation plots for a periodic process show multiple peaks emphasizing the occurrence of repeated events at a particular interval. Similar observations are made in the present case from Fig. \ref{fig:4.14}. The peaks corresponding to cross-correlation (shown in green color) do not occur at zero time lag. This implies that the periodic process is happening at both locations at a slight time delay. Physically, it suggests a flapping motion of the impinging shear layer such that the feedback waves reach at the nozzle exit one after the other. Also, it would mean that the impinging tones are produced due to the flapping motion of the shear layer for all the cases. This idea is further explored in the next section. The auto-correlation plots (shown in red color) are obtained by pressure values probed at point A. The peaks for this plot occur after nearly equal intervals of time. The time lag between each peak is reported in Tab. \ref{table:4.5}. This time lag corresponds to the dominant impinging tone frequency for each case. This observation proves that the impinging tones are indeed produced due to the loop closed by feedback waves moving upstream outside the jet.

Streamwise velocity values are probed at points C, D, E, F to estimate the convection velocity of large-scale structures. It is observed that the convective velocities have larger values near the nozzle exit and smaller values far from it. An average convective velocity is obtained for each case and presented in Tab. \ref{table:4.5}. These values also match with the average convective velocities reported in the literature.  

\begin{table}
	\caption{Skewness and Kurtosis of the pressure signal obtained at $x=0,y=8.5D$}
	\begin{tabularx}{\linewidth}{|X|X|X|}
		\hline
		\textbf{Case}&\textbf{Skewness}&\textbf{Kurtosis} \\
		\hline
		x/D-3pt94 & 0.24 & 2.86  \\
		\hline 
		x/D-5pt5 & 0.60 & 4.2 \\
		\hline 
		x/D-8pt27 & 0.15 & 2.98 \\
		\hline 
		x/D-9pt1 & 0.21 & 2.86\\
		\hline 
		x/D-5pt5-75deg & 0.23 & 2.96 \\
		\hline 
	\end{tabularx}
	\label{table:4.4}
\end{table}

\subsubsection{Far-Field Characteristics}

\begin{figure}
	\includegraphics[scale=0.5]{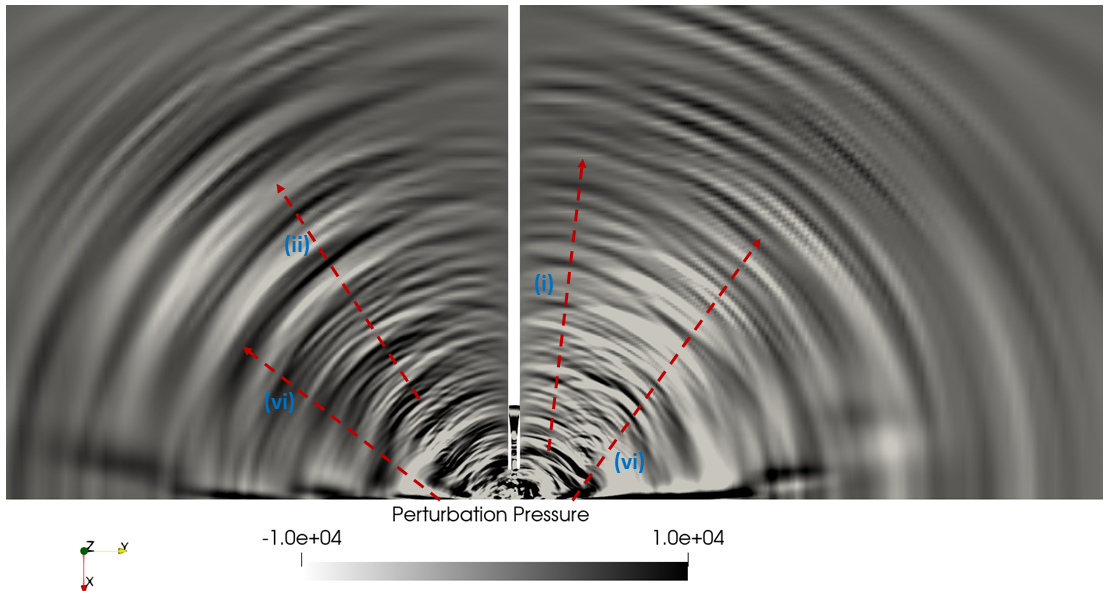}
	\centering
	\caption{\label{fig:4.15}Far-field propagation}	
\end{figure}

\begin{figure}
	\includegraphics[scale=0.6]{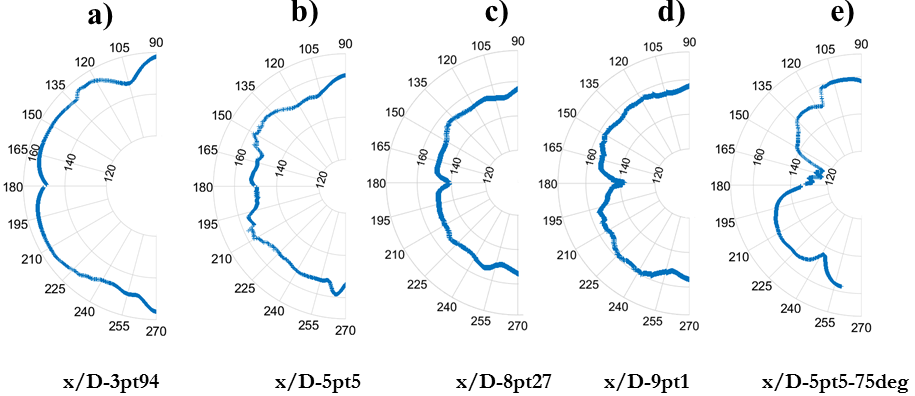}
	\centering
	\caption{\label{fig:4.16}OASPL at $100D$ from the jet exit:a) x/D-3pt94, b) x/D-5pt5, c) x/D-8pt27, d) x/D-9pt1, e) x/D-5pt5-75deg}	
\end{figure}

In the previous sections, the near-field flow features as well as acoustic signatures are discussed in detail. These features are also found in the far-field acoustic signature presented for x/D-5pt5 in Fig. \ref{fig:4.15}. The perturbation pressure contours reveal typical sound field of tonal noise filled with broadband noise emanating from various sources. The components represented with arrows marked as (i) and (ii) were identified previously. However, another component is identified in the present figure, marked as arrow (vi). This component is quite far away from the impingement region (more than 10D from the nozzle centerline). It has been suggested by Nonomura et al.\cite{nonomura2011aeroacoustic} that the Mach waves created in the wall jet region can also be the source of the acoustic waves. The peak velocity obtained previously in the wall jet was roughly 0.8-0.9 times the jet exit velocity. The convective velocity of large-scale structures would be further less in the wall jet. Therefore, the possibility of Mach wave radiation in the wall jet for the present case is slim. The source term at this point is possibly due to the spreading of the wall jet plume perpendicular to the plate. OASPL at $100D$ from the jet axis is presented for different cases in Fig. \ref{fig:4.16}. It is observed that for smaller plate distances ( Fig. \ref{fig:4.16}(a) and  Fig. \ref{fig:4.16}(b)), the OASPL is higher near the jet axis as compared to the larger plate distance (Fig. \ref{fig:4.16}(c) and  Fig. \ref{fig:4.16}(d)). However, the minimum OASPL near the axis is found to be for x/D-5pt5-75deg. This is consistent with the near-field observation. The effect of tonal noise for x/D-5pt5-75deg is higher along the positive $y-axis$ while along the negative $y-axis$, the OASPL shows a lack of major peaks and, therefore, the dominance of broadband noise. The tonal noise and the broadband components are equally distributed for other cases, as evident by the symmetric nature of contours across $x-axis$. It is interesting to note that the OASPL is relatively higher near the plate for all the cases. This can again be attributed to the spread of jet plume. 

\begin{table}
	\caption{Values obtained from correlation analysis}
	\begin{tabularx}{\linewidth}{|X|X|X|}
		\hline
		\textbf{Case}&\textbf{Peak correlation lag} (s)&\textbf{Average Convective Velocity} \\
		\hline
		x/D-3pt94 & 1.20e-4 & $0.60U_j$ \\
		\hline 
		x/D-5pt5 & 1.60e-4 & $0.67U_j$ \\
		\hline 
		x/D-8pt27 & 2.90e-4 & $0.69U_j$ \\
		\hline 
		x/D-9pt1 & 3.34e-4 & $0.72U_j$\\
		\hline 
		x/D-5pt5-75deg & 1.15e-4 & $0.66U_j$ \\
		\hline 
	\end{tabularx}
	\label{table:4.5}
\end{table}

\subsection{Modal Decomposition}

A key feature of the jets is the existence of different oscillation modes. These oscillation modes are associated with the change in mode number for impinging jets \cite{henderson1993experiments} and can, thereby, change the frequency of the impinging tone. Since the present work employs a periodic condition at the spanwise plane, the present configuration is close to a rectangular jet with a high aspect ratio. The only oscillations which are taken into consideration, consequently, are the anti-symmetric oscillations about the spanwise axis (also called sinuous mode) and the symmetric oscillations (also called varicose mode). The presence of these modes is ascertained with the help of Proper Orthogonal Decomposition (POD) and Dynamic Mode Decomposition (DMD).  


\subsubsection{Proper Orthogonal Decomposition (POD)}

\begin{figure}
	\includegraphics[scale=0.2]{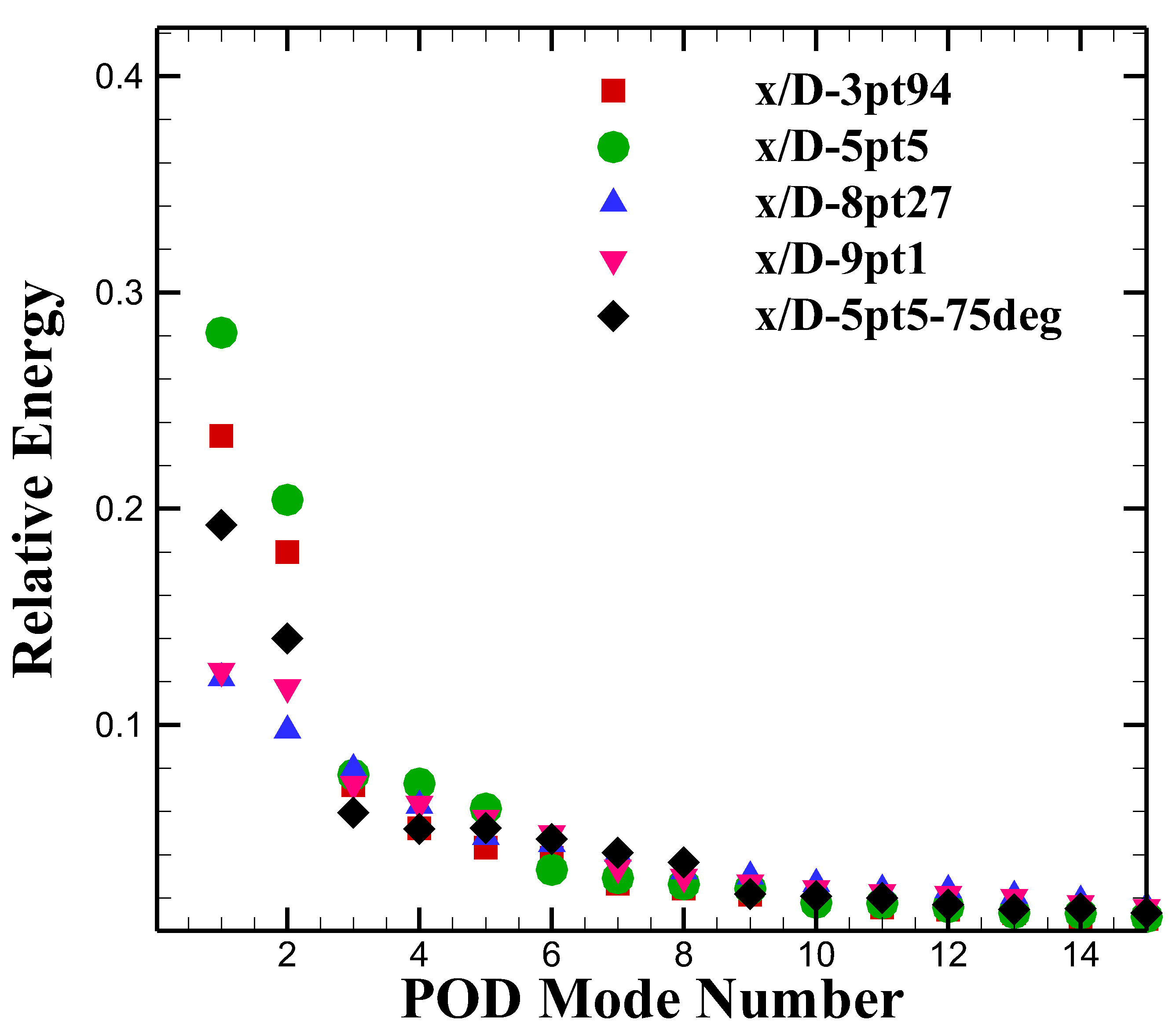}
	\centering
	\caption{\label{fig:4.17}Relative Eignevalue of different modes for all the cases}	
\end{figure}

\begin{figure}
	\includegraphics[scale=0.2]{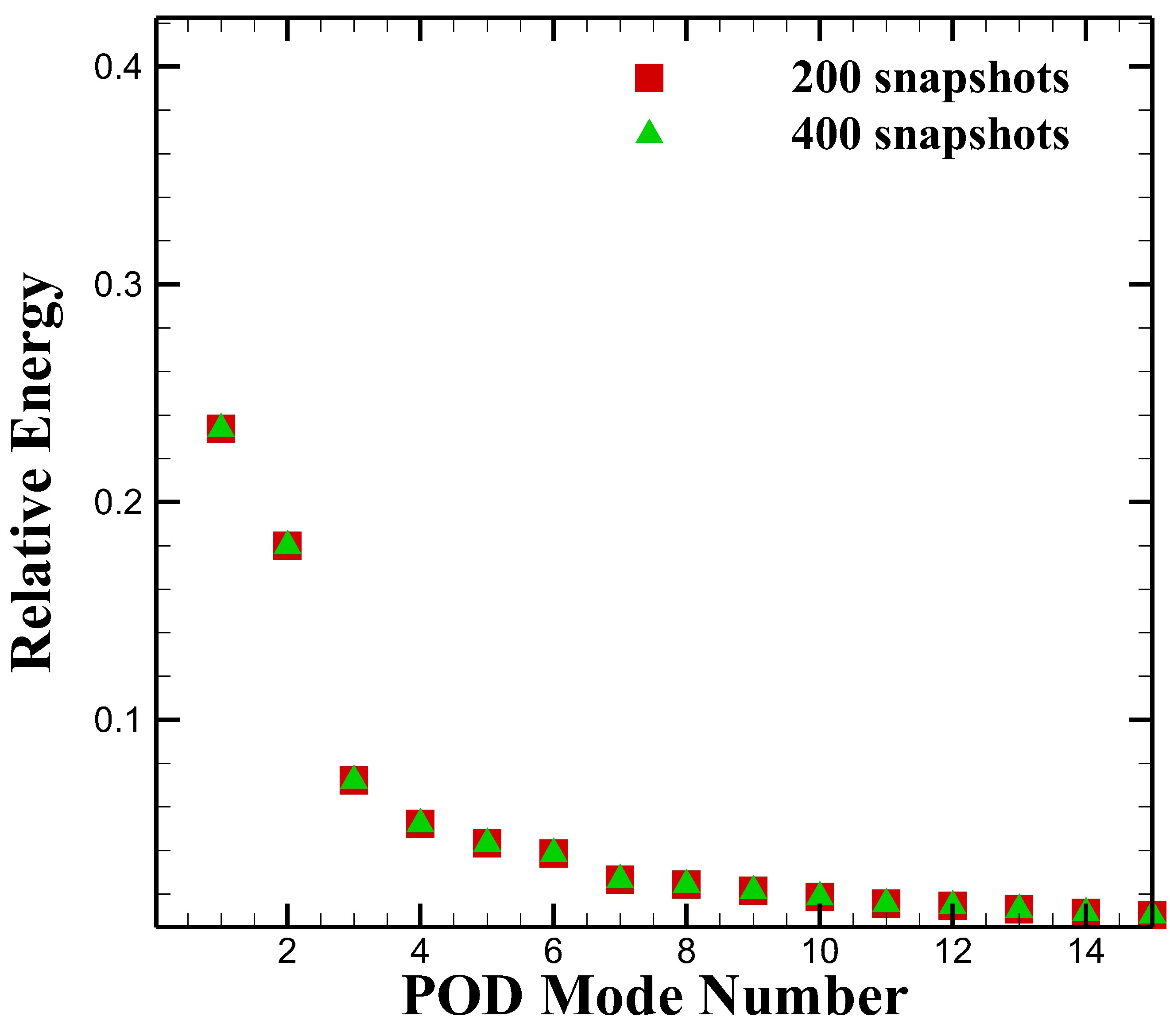}
	\centering
	\caption{\label{fig:4.32}Relative Eigenvalues for different number of Snapshots for x/D-3pt94}	
\end{figure}

POD is a mathematical tool to convert a series of observations into spatio-temporal modes such that the important properties can be defined by a few modes. In the context of turbulent flows, POD was introduced by Lumley \cite{lumley1967structure}. Various POD modes define the coherent structures present in any turbulent flow. We perform an energy-based POD using the method of snapshots \cite{sirovich1987turbulence} in which an auto-covariance matrix is created from the velocity components. For compressible flows, the Temperature fluctuations are also included in the auto-covariance matrix. Therefore, a scaling parameter \cite{lumley1997low} is required to incorporate the effect of Temperature change on the total energy. The details of the implementation of this methodology can be found out in Soni et al \cite{soni2017characterization}.  Initially, the POD procedure is performed using 200 snapshots in time, including 20 periods of oscillations for the dominant frequency in each case. The eigenvalues obtained from the decomposition for more than 200 snapshots are nearly the same, as evidenced from Fig. \ref{fig:4.32}. Therefore, the POD analysis is presented using 200 snapshots in time. The relative eigenvalue, which is a measure of the energy content of a mode, is presented in Fig. \ref{fig:4.17}. It is observed that the first few modes in each case are associated with maximum energy. The important dynamics of the flow can be observed by studying the properties of the eigenvectors corresponding to the maximum eigenvalues. Different features of the flow are captured in modes corresponding to velocity and Temperature. The first mode corresponding to velocity along the $x-direction$ is written as $\phi_{1x}$, the first mode corresponding to Temperature is written as $\phi_{1T}$, and so on. 

\begin{figure}
	\includegraphics[scale=0.45]{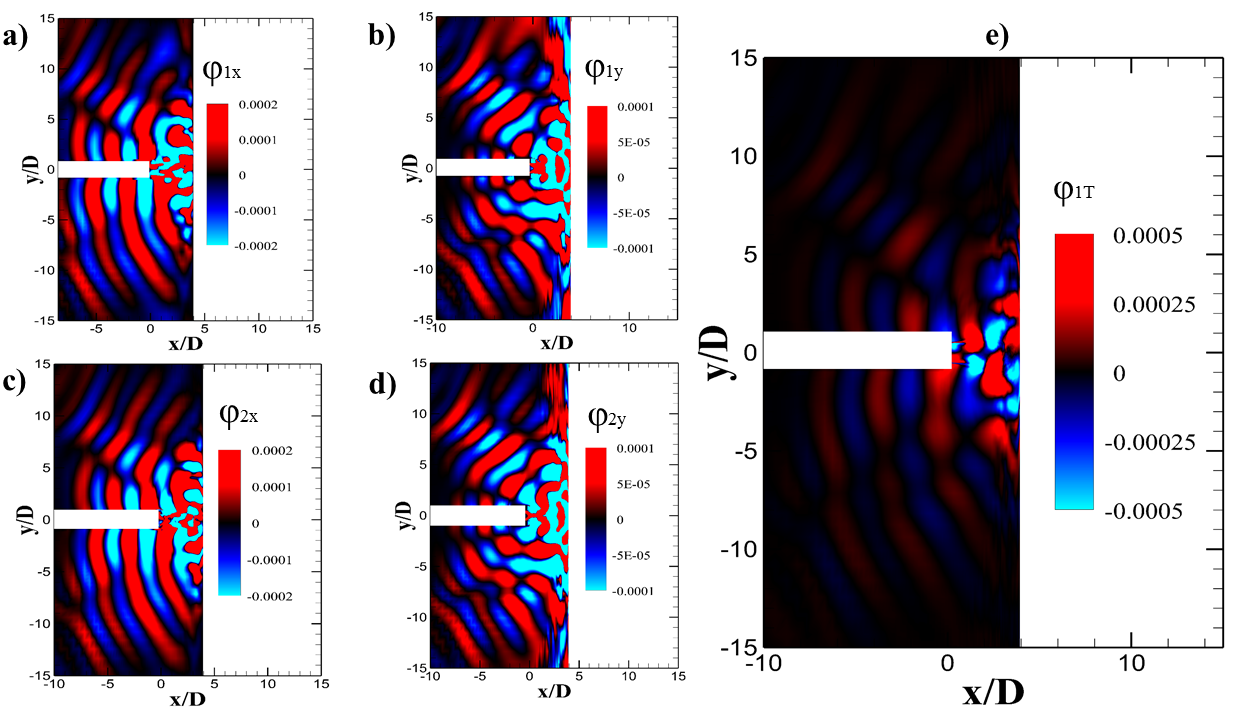}
	\centering
	\caption{\label{fig:4.18}POD modes for x/D-3pt94}	
\end{figure}
\begin{figure}
	\includegraphics[scale=0.45]{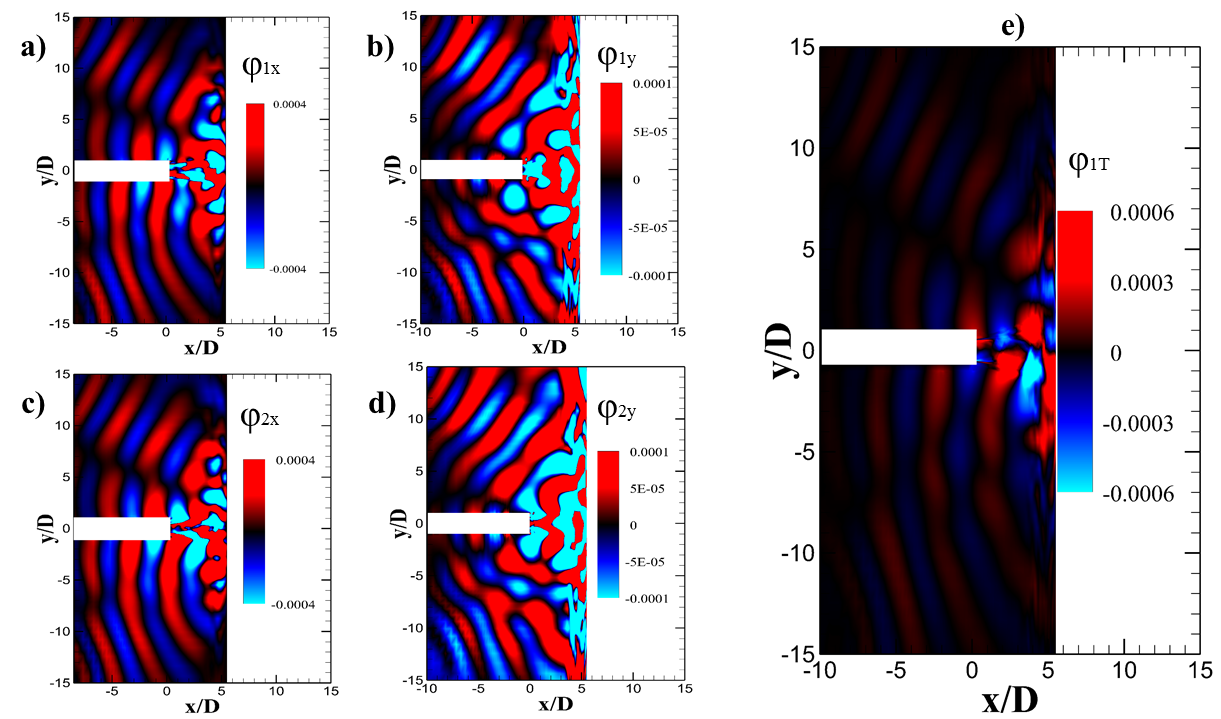}
	\centering
	\caption{\label{fig:4.19}POD modes for x/D-5pt5}	
\end{figure}

\begin{figure}
	\includegraphics[scale=0.45]{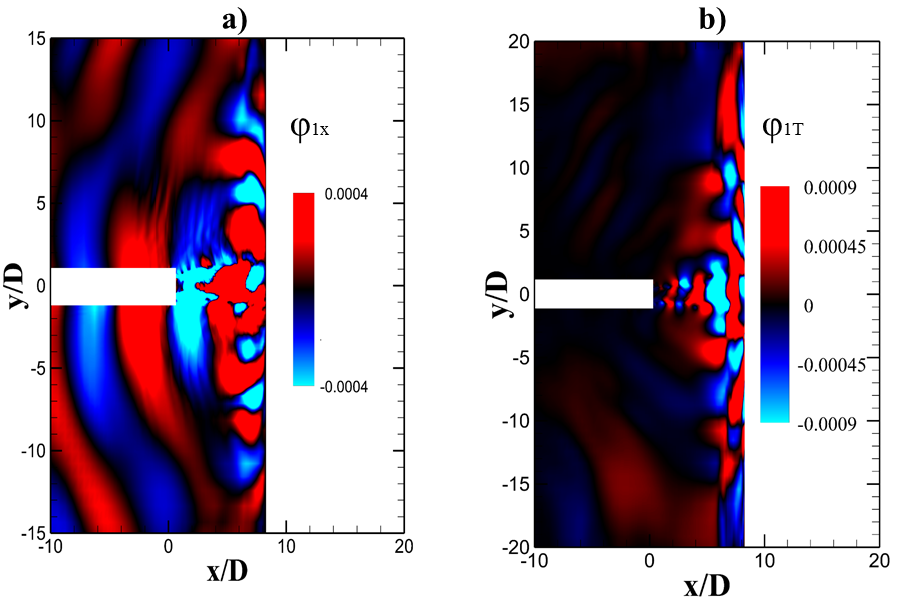}
	\centering
	\caption{\label{fig:4.20}POD modes for x/D-8pt27}	
\end{figure}

\begin{figure}
	\includegraphics[scale=0.45]{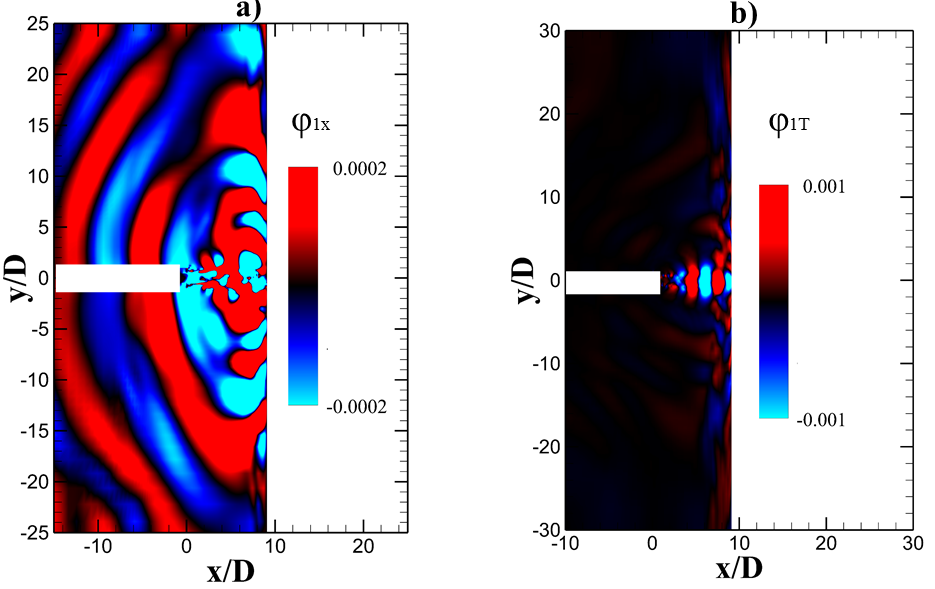}
	\centering
	\caption{\label{fig:4.21}POD modes for x/D-9pt1}	
\end{figure}

In Fig. \ref{fig:4.18} ((a)-(d)), the first two modes corresponding to $x-velocity$ and $y-velocity$ and the first mode corresponding to Temperature are presented for the case x/D-3pt94. We observe that the POD modes along $x-velocity$ are anti-symmetric with respect to the jet axis, whereas the POD modes along $y-velocity$ are symmetric with respect to the jet axis. This same observation can be made from Fig. \ref{fig:4.19}((a)-(d)) for x/D-5pt5. The anti-symmetric and symmetric structures correspond to the sinuous and varicose modes, respectively. Since the first two modes correspond to the largest energy, both modes are equally likely for smaller plate distances. The structure obtained in the Temperature mode for both the cases (Fig. \ref{fig:4.18}(e) and Fig. \ref{fig:4.19}(e)) is composed of three pairs of lobes for each case. This number can be related to the mode number, $N$, in Powell's Equation (Eq. 1). It will be observed in the next section that the mode number for x/D-3pt94 and x/D-5pt5 is one more than the number of pairs of lobes obtained. So, it means that the mode number, $N$, for the tonal frequency for smaller plate distances is 4.     
For the larger plate distance, the first mode along $x-velocity$ (Fig. \ref{fig:4.20}(a) and Fig. \ref{fig:4.21}(a))  and the first mode corresponding to Temperature (Fig. \ref{fig:4.20}(b) and Fig. \ref{fig:4.21}(b)) is presented. The other modes are similar to these modes and therefore not presented here. It is observed that the structure for the mode along $x-velocity$ is symmetric about the jet axis, and therefore it represents symmetric oscillation mode. Therefore, for larger plate distances, symmetric oscillations are associated with higher energy. The mode corresponding to Temperature in each case represents 4 lobes. This is again related to the mode number for the impinging tones. For larger plate distances, the mode number is one less than the number of lobes. Thus, for larger plate distances, the mode number, $N$, is 3.

\begin{figure}
	\includegraphics[scale=0.45]{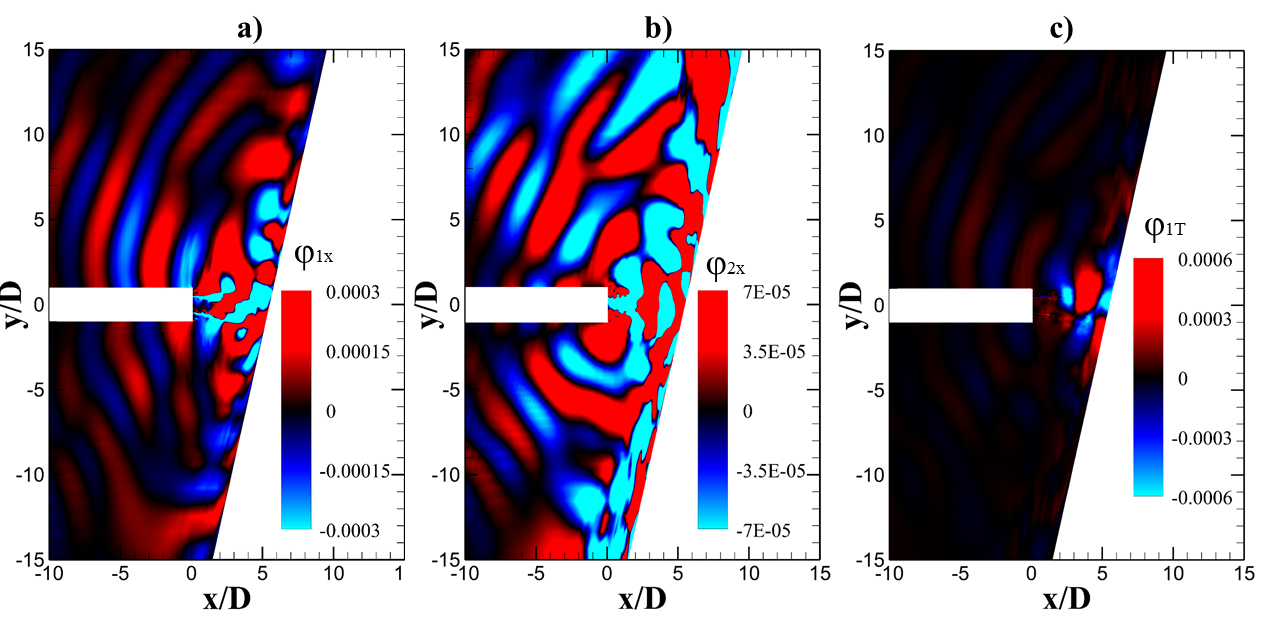}
	\centering
	\caption{\label{fig:4.22}POD modes for x/D-5pt5-75deg}	
\end{figure}

The POD modes for x/D-5pt5-75deg presented in Fig. \ref{fig:4.22} display a predominance of anti-symmetric oscillations. The mode number, $N$, in this case is also 4.

Though the POD modes provide us valuable information of different oscillation modes in the present case, it is difficult to relate them with the impinging tone frequencies. This is because the POD modes are segregated based on their energy content. Any POD mode will consist of multiple frequencies whose information can be extracted from the temporal coefficients. However, the mode shapes corresponding to these frequencies cannot be predicted by POD. This difficulty can be overcome by employing a Dynamic Mode Decomposition (DMD) to the current dataset.

\subsubsection{Dynamic Mode Decomposition (DMD)} 

DMD is a data analysis technique \cite{schmid2011applications} that finds a best fit linear model for advancing any measurement at a particular time, into a next time step. The key advantage is that the underlying data can be an output of a linear or a non-linear system. Since the DMD is a linear approximation of the future state of any variable based on a present observation, it is helpful in extracting dynamical information from the sampled data. Each DMD mode is composed of a single frequency. In the case of fluids, the DMD modes which are neither damped nor amplified represent the most important modes. Such modes provide useful insights into the flow. A detailed description of the DMD methodology and its implementation can be found in Soni et al \cite{soni2019modal}. The number of snapshots used for the DMD is equal to 200. The time difference between these snapshots ensures that the Nyquist criterion is satisfied for the correct resolution of frequencies \cite{soni2019modal}. For the present case, the real part of the DMD modes is presented for all cases.

\begin{figure}
	\begin{subfigure}{1\textwidth}
		\includegraphics[scale=0.5]{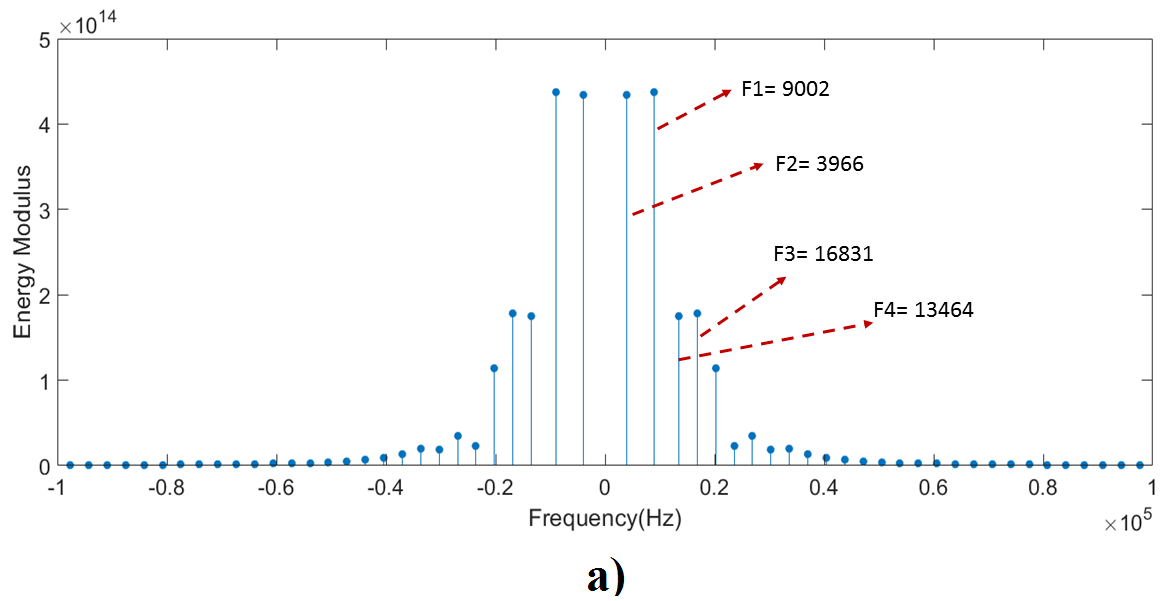}  
		\centering
	\end{subfigure}
	
	\begin{subfigure}{1\textwidth}
		\includegraphics[scale=0.48]{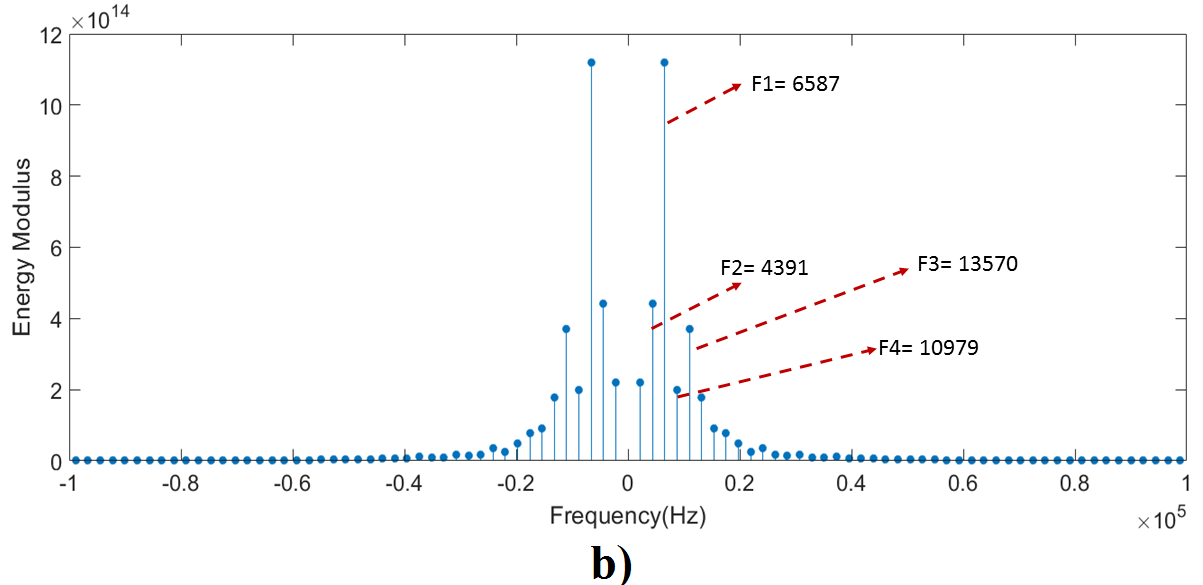}  
		\centering
	\end{subfigure}	
	\begin{subfigure}{1\textwidth}
		\includegraphics[scale=0.5]{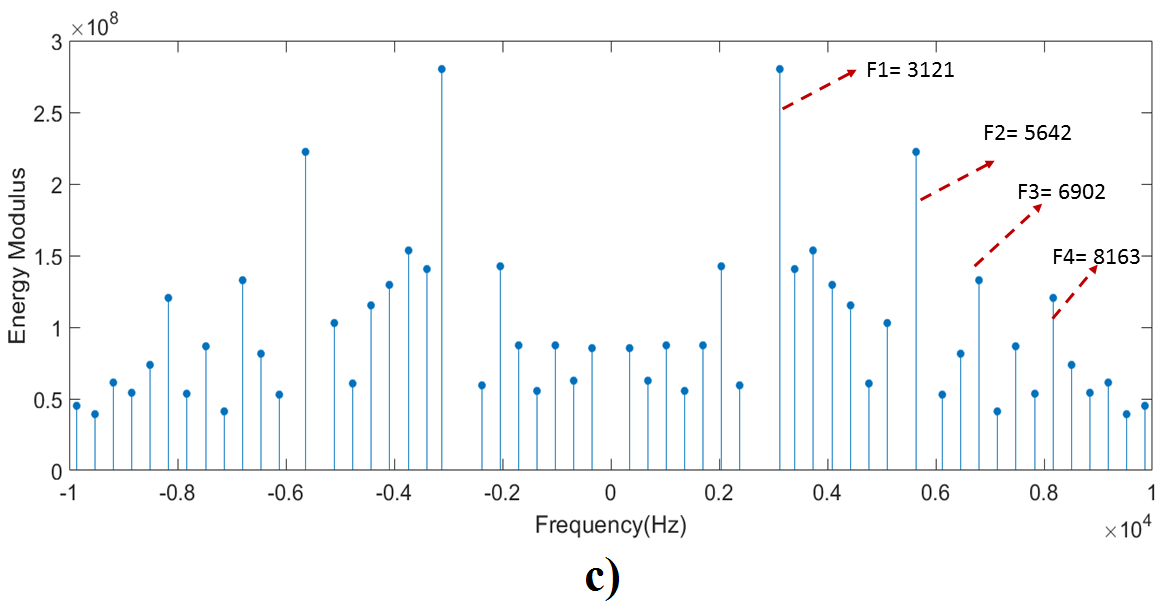}  
		\centering
	\end{subfigure}
\end{figure}
\begin{figure}
	\ContinuedFloat
	\begin{subfigure}{1\textwidth}
		\includegraphics[scale=0.5]{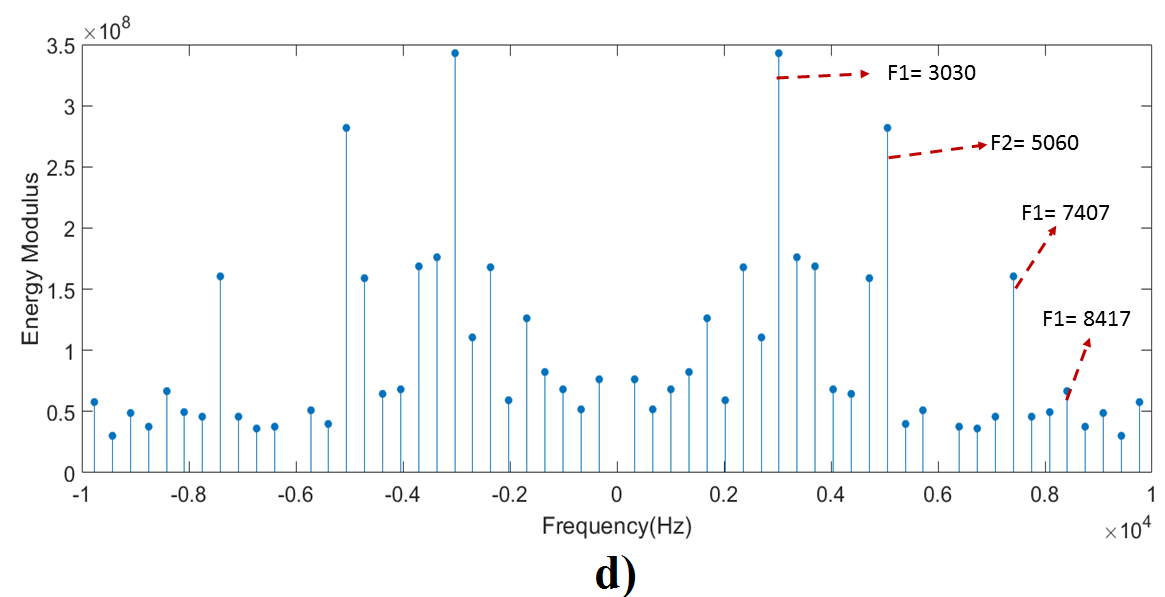}  
		\centering
	\end{subfigure}
	
	\begin{subfigure}{1\textwidth}
		\includegraphics[scale=0.5]{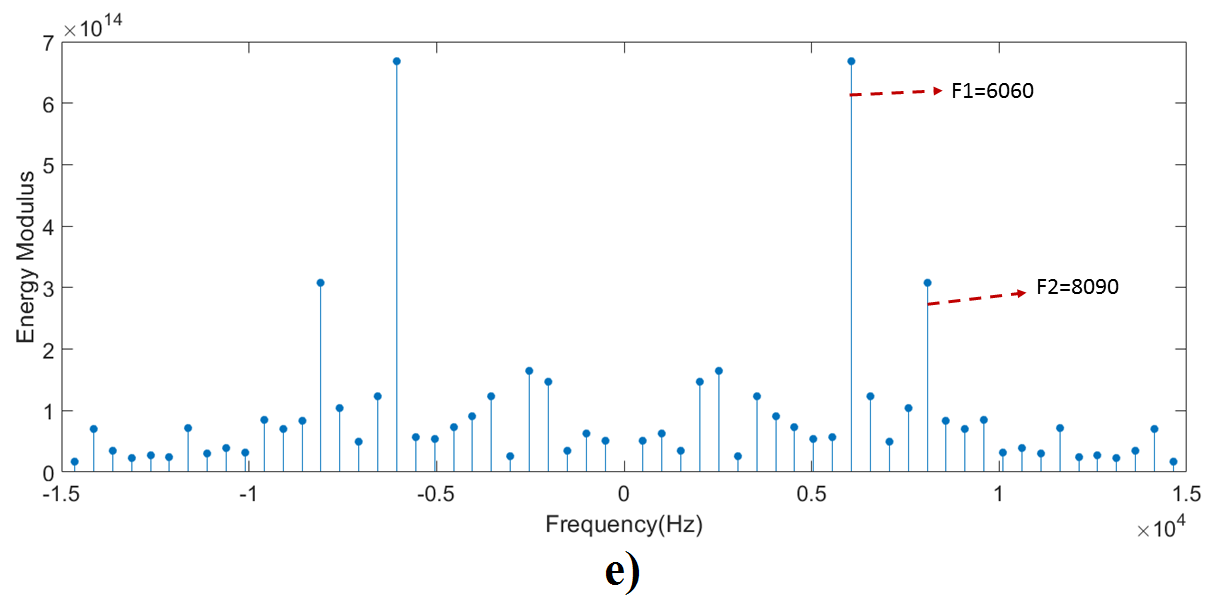}  
		\centering
	\end{subfigure}
	
	\caption{DMD Spectrum for different cases: a) x/D-3pt94, b) x/D-5pt5, c) x/D-8pt27, d) x/D-9pt1, e) x/D-5pt5-75deg}
	\label{fig:4.23}
\end{figure}

\begin{figure}
	\includegraphics[scale=0.6]{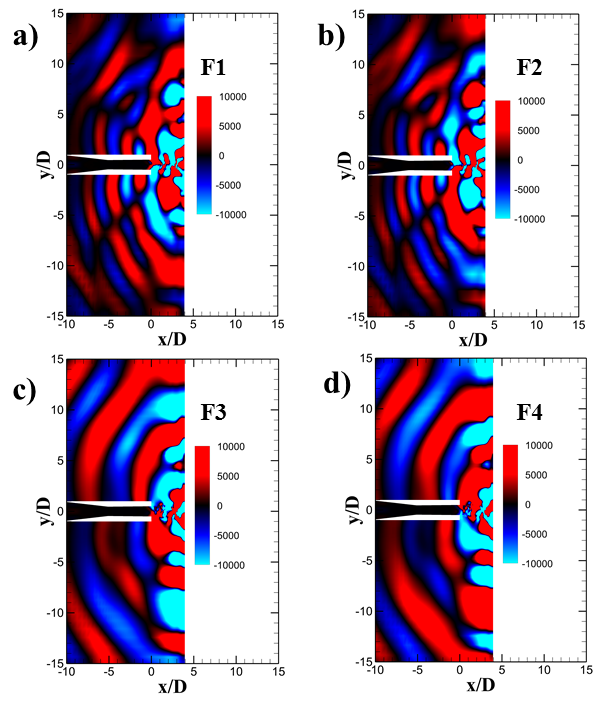}
	\centering
	\caption{\label{fig:4.24}DMD modes for x/D-3pt94}	
\end{figure}

\begin{figure}
	\includegraphics[scale=0.6]{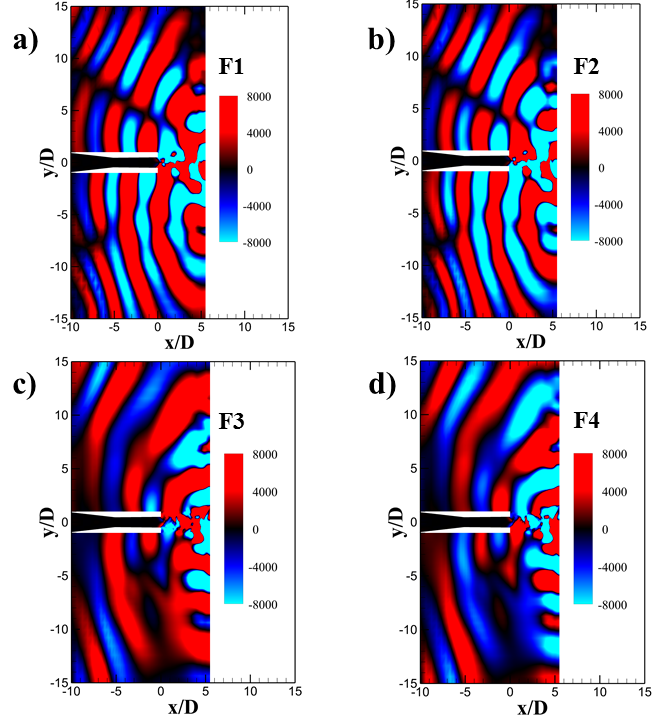}
	\centering
	\caption{\label{fig:4.25}DMD modes for x/D-5pt5}	
\end{figure}

\begin{figure}
	\includegraphics[scale=0.6]{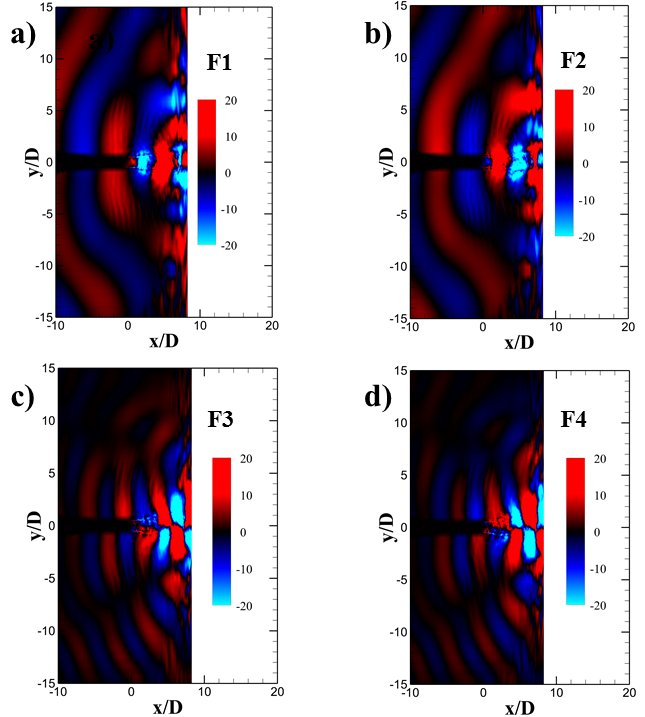}
	\centering
	\caption{\label{fig:4.26}DMD modes for x/D-8pt27}	
\end{figure}

\begin{figure}
	\includegraphics[scale=0.6]{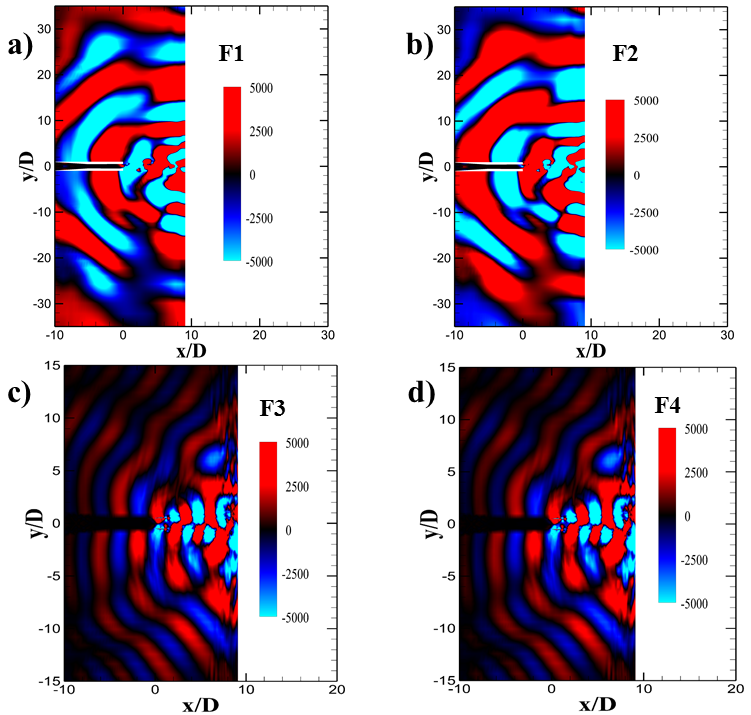}
	\centering
	\caption{\label{fig:4.27}DMD modes for x/D-9pt1}	
\end{figure}

\begin{figure}
	\includegraphics[scale=0.5]{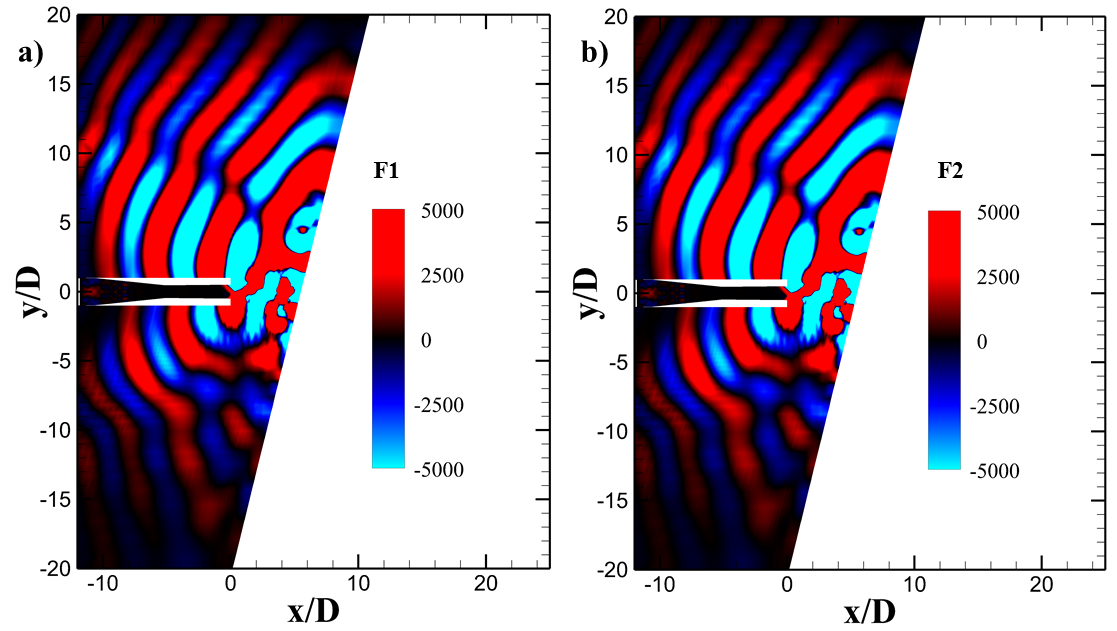}
	\centering
	\caption{\label{fig:4.28}DMD modes for x/D-5pt5-75deg}	
\end{figure}

The spectra obtained from DMD are presented in Fig. \ref{fig:4.23}. The frequencies captured in the DMD spectra and also reported in the reference study are highlighted in Fig. \ref{fig:4.23}. The mode shapes corresponding to the marked frequencies are presented for different cases. It is observed in Fig. \ref{fig:4.24} that mode shapes corresponding to the first four frequencies for x/D-3pt94 are anti-symmetric. For x/D-5pt5, as seen in Fig. \ref{fig:4.25}, the first two frequencies are anti-symmetric. However, the higher frequencies do not show any definite pattern. It leads us to conclude that for smaller plate distances, the dominant impinging tones are produced by anti-symmetric oscillations. It is observed from Fig. \ref{fig:4.26} and Fig. \ref{fig:4.27} (larger plate distance) that the first two frequencies correspond to a symmetric mode while the higher frequencies correspond to an anti-symmetric mode of oscillation. The two dominant frequencies for x/D-5pt5-75deg in Fig. \ref{fig:4.28} display an anti-symmetric mode of oscillations.

The existence of different modes of oscillations for different harmonics of screech tones in rectangular jets had been reported by Raman and Rice \cite{raman1994instability}. However, the existence of different modes in a particular case for different tones generated due to the supersonic jet impingement process is scarcely reported. The present analysis confirms the presence of different modes of oscillations for different impinging tones for a particular nozzle to plate distance.

\begin{figure}
	\includegraphics[scale=0.55]{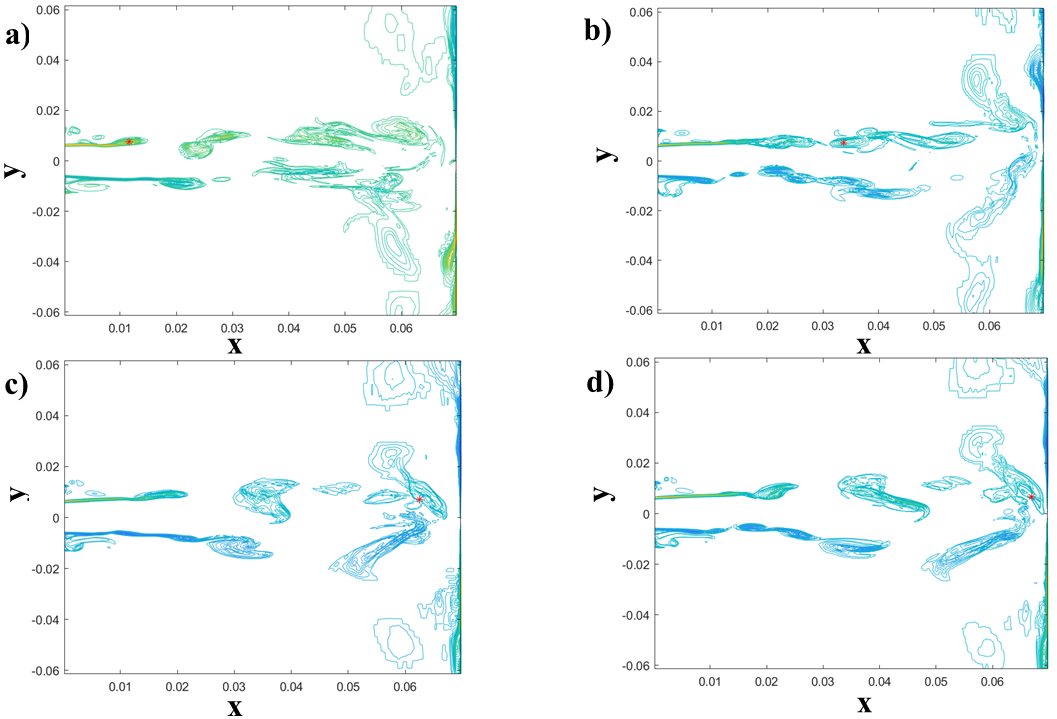}
	\centering
	\caption{\label{fig:4.29}Vortex tracking showing impinging vortices}	
\end{figure}

\subsection{Powell's Equation Revisited using Vortex Tracking}

We now present a quantitative estimate of the frequencies present in the system by calculating the time taken by vortices to reach the plate by a vortex tracking method. The vortex centers are identified by regions of high vorticity in the domain. The process first includes the detection of a vortex center as it is shed from the nozzle exit. In the next step, high vorticity points are located in a small region, usually downstream of the initial point. This would locate the vortex as it is convected downstream. The vortex detection algorithm sometimes detects the center of a different vortex which is interacting with the currently tracked vortex center. Ad-hoc arrangements are made to account for such cases.

\begin{figure}
	\includegraphics[scale=0.55]{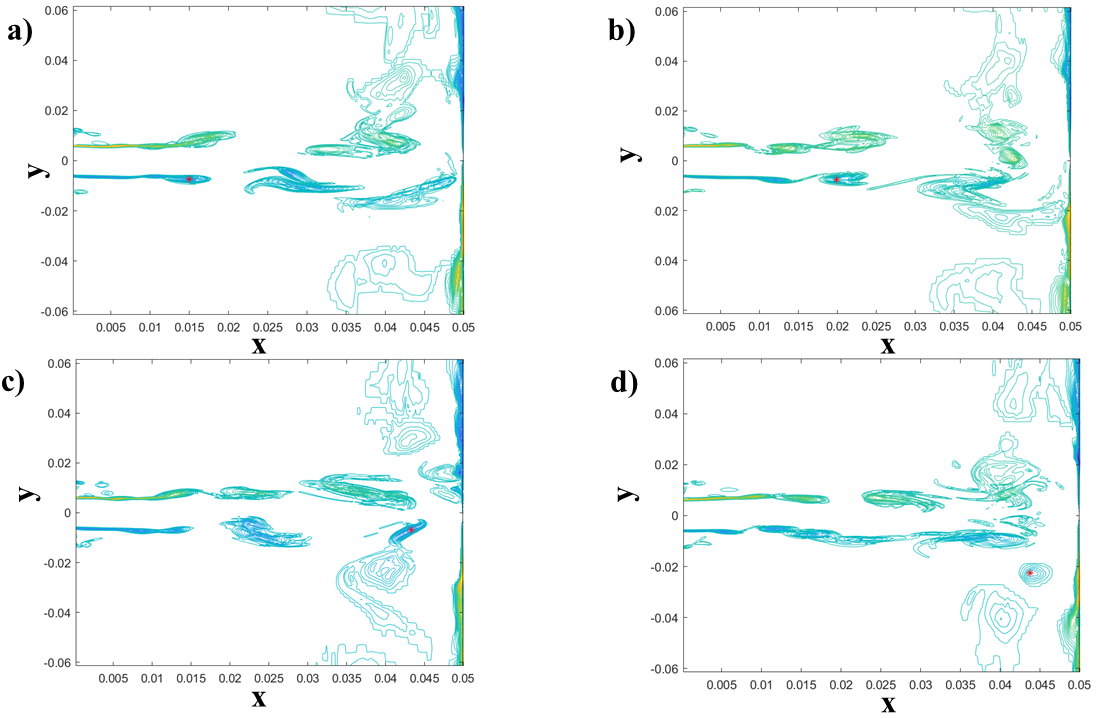}
	\centering
	\caption{\label{fig:4.30}vortex tracking showing vortex movement along the wall jet}	
\end{figure}
\begin{figure}
	\includegraphics[scale=0.2]{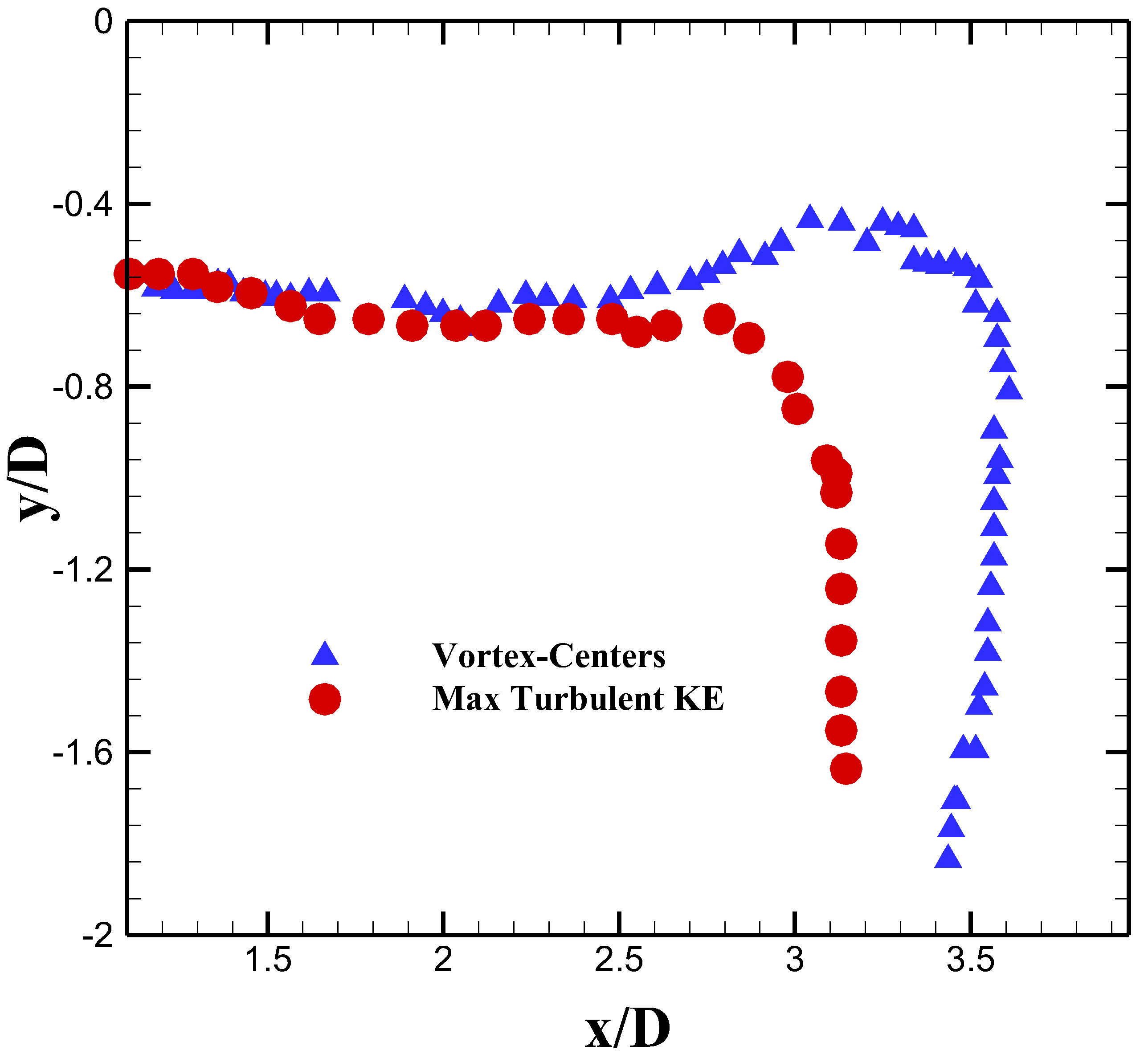}
	\caption{\label{fig:4.31}vortex centers and location of maximum turbulent kinetic energy}	
\end{figure}

It is found that as the vortices are convected downstream, there are two major paths they can follow. In the first case, they directly impinge on the wall. It is represented by Fig. \ref{fig:4.29}. The impingement points for the cases considered presently lie between $0.2D$ to $1.5D$. In the second case, instead of directly impinging on the wall, they are convected into the wall jet. This case is represented by Fig. \ref{fig:4.30}. In this case, the vortex slightly moves in the upstream direction before going into the wall jet. This upstream movement roughly begins at a distance of $1.35D$ from the jet centerline. This is evident from Fig. \ref{fig:4.31}. The upstream travel of vortices and subsequent movement into the wall jet leads to the pulsation of the wall jet as noticed by Henderson et al\cite{henderson2005experimental}. The location of the maximum turbulent kinetic energy is also presented in Fig. \ref{fig:4.31}. It has been suggested by Weightman et al. \cite{weightman2017explanation} that the location of maximum turbulent kinetic energy can be used as an approximate path of the vortices. As observed, the vortex path and the maximum turbulent kinetic energy location coincide in the initial shear layer. However, near the wall jet region where complex interactions happen, there is a deviation. 

The vortex detection algorithm calculates the time taken by vortices to reach the impinging point. The feedback waves travel at the speed of the sound. The average speed of travel, however, is slightly less due to the entrainment effects. A value of $331$ m/s is used according to Weightman et al \cite{weightman2017explanation}. The mode number, $N$, for each case had been discussed in the POD analysis. These values are fed to Powell's equation, and the value of phase lag term corresponding to the dominant impinging tone is estimated. The values of the phase lag term obtained are presented in Tab. \ref{table:4.6}. We observe that the value obtained is nearly $0.41$ which is fairly close to the value used by Krothapalli \cite{krothapalli1999flow} to match the frequencies obtained from his experiments. If we assume a time delay associated with the impingement of large scale structures and generation of feedback waves, we can see that the value of this time delay corresponding to the constant phase lag term increases for increasing plate distance. This observation is not trivial. If the dominant tone is caused due to the feedback wave generated by the shocklet movement, it would mean that the shocklet movement is dependent on the plate distance. There could also be a different process altogether to corroborate this observation which is a subject of a future study.

\begin{table}
	\caption{Phase value $p$ for each case}
	\begin{tabularx}{\linewidth}{|X|X|X|X|}
		\hline
		\textbf{Case}&\textbf{N}&\textbf{p}&\textbf{$\Delta t(\mu s)$} \\
		\hline
		x/D-3pt94 & 4 & -0.41 &46 \\
		\hline 
		x/D-5pt5 & 4 & -0.42 & 62 \\
		\hline 
		x/D-8pt27 & 3 & -0.40 & 122 \\
		\hline 
		x/D-9pt1 & 3 & -0.41 & 136 \\
		\hline 
		x/D-5pt5-75deg & 4 & -0.42 & 62\\
		\hline 
	\end{tabularx}
	\label{table:4.6}
\end{table}

\section{\label{sec:level1}Conclusion}   

We study the near and far-field noise for a planar supersonic impinging jet in this work by employing an LES along with a hybrid approach for the solution of far-field acoustic pressure. The averaged fluctuations are compared with a reference study and we find that they are in a reasonable agreement. From the near-field analysis, we find that the tonal components propagate travel upstream at small angles with respect to the jet axis. Mach wave radiation for the present cases is not significant, however, broadband noise is associated with the jet shear layer. In addition to these sources, a new acoustic source is identified which is located nearly at $y=10D$ from the jet centerline. This source corresponds to the spreading of the jet plume in the upstream direction. The far-field noise shows signature of tonal as well as broadband components. The OASPL for perpendicular impingement is symmetric whereas for angled impingement, tonal noise is focused along the positive $y-axis$ while broadband noise is dominant along negative $y-axis$. POD and DMD are used to identify the nature of jet oscillations during impingement. It is found that for smaller plate distance, symmetric as well as anti-symmetric oscillations are predominant while for larger plate distance, symmetric mode of oscillation is dominant. For smaller plate distance, the dominant impinging tones correspond to anti-symmetric oscillations whereas for larger plate distance, low-frequency tones correspond to symmetric mode of oscillations and high-frequency tones correspond to anti-symmetric oscillations. The vortex detection algorithm reveals the evolution of the vortices as they are shed from the nozzle exit and reach the plate. The time taken by these vortices is calculated and a phase lag term is obtained using Powell's Equation. The phase lag term is found to be nearly constant for all the cases. The time delay happening at the impingement region, therefore, increases for increasing plate distance. The reason and the exact mechanism of this process is a subject of future study.

\section*{Data Availability}

The data that support the findings of this study are available from the corresponding author upon reasonable request.

\section*{Acknowledgement}
Financial support for this research is provided through IITK-Space Technology Cell (STC). The authors would also like to acknowledge the High-Performance Computing (HPC) Facility at IIT Kanpur (www.iitk.ac.in/cc).

\section*{References}

\bibliography{aipsamp}
\clearpage

\end{document}